\documentclass[twocolumn,tighten]{aastex631}

\usepackage{float}
\usepackage[caption = false]{subfig}

\newcommand{\alphaforsec}{\texorpdfstring{$\alpha$}{alpha}}
\newcommand{\hiiforsec}{\texorpdfstring{\ion{H}{2}}{HII}}

\newcommand{\OSU}{\affil{Department of Astronomy, The Ohio State University, 140 West 18th Avenue, Columbus, Ohio 43210, USA}}

\newcommand{\CCAPP}{\affil{Center for Cosmology and Astroparticle Physics, 191 West Woodruff Avenue, Columbus, OH 43210, USA}}

\newcommand{\COOL}{\affil{Cosmic Origins Of Life (COOL) Research DAO, coolresearch.io}}

\newcommand{\ESO}{\affil{European Southern Observatory, Karl-Schwarzschild Stra{\ss}e 2, D-85748 Garching bei M\"{u}nchen, Germany}}

\newcommand{\ITA}{\affiliation{Universit\"{a}t Heidelberg, Zentrum f\"{u}r Astronomie, Institut f\"{u}r Theoretische Astrophysik, Albert-Ueberle-Str 2, D-69120 Heidelberg, Germany}}

\newcommand{\IWR}{\affiliation{Universit\"{a}t Heidelberg, Interdisziplin\"{a}res Zentrum f\"{u}r Wissenschaftliches Rechnen, Im Neuenheimer Feld 205, D-69120 Heidelberg, Germany}}

\newcommand{\NRAO}{\affil{National Radio Astronomy Observatory, 520 Edgemont Road, Charlottesville, VA 22903-2475, USA}}

\newcommand{\stsci}{\affil{Space Telescope Science Institute, Baltimore, MD USA}}
\newcommand{\UWyoming}{\affiliation{Department of Physics and Astronomy, University of Wyoming, Laramie, WY 82071, USA}}

\newcommand{\MPIA}{\affil{Max-Planck-Institut f\"{u}r Astronomie, K\"{o}nigstuhl 17, D-69117, Heidelberg, Germany}}
\newcommand{\ANU}{\affil{Research School of Astronomy and Astrophysics, Australian National University, Canberra, ACT 2611, Australia}}

\newcommand{\JHU}{\affil{Center for Astrophysical Sciences, Johns Hopkins University, Baltimore, MD 21218, USA}}

\newcommand{\oxford}{\affil{Sub-department of Astrophysics, Department of Physics, University of Oxford, Keble Road, Oxford OX1 3RH, UK}}

\newcommand{\ARI}{\affil{Astronomisches Rechen-Institut, Zentrum f\"{u}r Astronomie der Universit\"{a}t Heidelberg, M\"{o}nchhofstra\ss e 12-14, D-69120 Heidelberg, Germany}}

\shorttitle{SNe near \ion{H}{2} regions}
\shortauthors{Mayker~Chen et al.}

\begin{document}

\title{H$\alpha$ emission and \ion{H}{2} regions at the locations of recent supernovae in nearby galaxies}

%------------------------ABSTRACT-----------------------------
\begin{abstract}
We present a statistical analysis of the local, $\approx 50{-}100$~pc scale, H$\alpha$ emission at the locations of recent ($\leq125$~years) supernovae (SNe) in nearby star-forming galaxies. Our sample consists of 32 SNe in 10 galaxies that are targets of the PHANGS-MUSE survey. We find that 41\% (13/32) of these SNe occur coincident with a previously identified \ion{H}{2} region. For comparison, \ion{H}{2} regions cover 32\% of the area within $\pm 1$~kpc of any recent SN. Contrasting this local covering fraction with the fraction of SNe coincident with \ion{H}{2} regions, we find a statistical excess of 7.6\% $\pm$ 8.7\% of all SNe to be associated with \ion{H}{2} regions. This increases to an excess of 19.2\% $\pm$ 10.4\% when considering only core-collapse SNe. These estimates appear to be in good agreement with qualitative results from new, higher resolution \textit{HST} H$\alpha$ imaging, which also suggest many CCSNe detonate near but not in \ion{H}{2} regions. Our results appear consistent with the expectation that only a modest fraction of stars explode during the first $\lesssim 5$~Myr of the life of a stellar population, when H$\alpha$ emission is expected to be bright. Of the \ion{H}{2} region associated SNe, 85\% (11/13) also have associated detected CO~(2--1) emission, indicating the presence of molecular gas. The \ion{H}{2} region associated SNe have typical extinctions of $A_V \sim1$~mag, consistent with a significant amount of pre-clearing of gas from the region before the SNe explode. 
\end{abstract}


%-----------AUTHOR INFORMATION------%

\correspondingauthor{Ness Mayker~Chen}
\email{maykerchen.1@osu.edu}

\author[0000-0002-5993-6685]{Ness Mayker~Chen}
\OSU \CCAPP

\author[0000-0002-2545-1700]{Adam~K.~Leroy}
\OSU\CCAPP

\author[0000-0002-4781-7291]{Sumit K. Sarbadhicary}
\OSU\CCAPP

\author[0000-0002-1790-3148]{Laura A. Lopez}
\OSU\CCAPP

\author[0000-0003-2377-9574]{Todd A. Thompson}
\OSU\CCAPP

\author[0000-0003-0410-4504]{Ashley T. Barnes}
\ESO

\author[0000-0002-6155-7166]{Eric Emsellem}
\affiliation{European Southern Observatory, Karl-Schwarzschild-Stra{\ss}e 2, 85748 Garching, Germany}
\affiliation{Univ Lyon, Univ Lyon1, ENS de Lyon, CNRS, Centre de Recherche Astrophysique de Lyon UMR5574, F-69230 Saint-Genis-Laval France}

\author[0000-0002-9768-0246]{Brent Groves}
\affiliation{International Centre for Radio Astronomy Research, University of Western Australia, 7 Fairway, Crawley, 6009 WA, Australia}

\author[0000-0003-0085-4623]{Rupali~Chandar}\affiliation{Ritter Astrophysical Research Center, University of Toledo, Toledo, OH 43606, USA}

\author[0000-0002-5635-5180]{M\'{e}lanie~Chevance}
\ITA
\COOL

\author[0000-0001-8241-7704]{Ryan Chown}
\OSU

\author[0000-0002-5782-9093]{Daniel~A.~Dale}
\UWyoming

\author[0000-0002-4755-118X]{Oleg V. Egorov}
\ARI

\author[0000-0001-6708-1317]{Simon~C.~O.~Glover}
\ITA

\author[0000-0002-3247-5321]{Kathryn~Grasha}
\ANU
\altaffiliation{ARC DECRA Fellow}

\author[0000-0002-0560-3172]{Ralf S.\ Klessen}
\ITA
\IWR

\author[0000-0001-6551-3091]{Kathryn Kreckel}
\ARI

\author[0000-0002-4825-9367]{Jing Li}
\ARI

\author[0000-0002-6972-6411]{J. Eduardo M\'endez-Delgado}
\ARI

\author[0000-0001-7089-7325]{Eric J. Murphy}
\NRAO

\author[0000-0003-2721-487X]{Debosmita~Pathak}
\OSU\CCAPP

\author[0000-0002-3933-7677]{Eva Schinnerer}
\MPIA

\author[0000-0002-8528-7340]{David A. Thilker}
\JHU

\author[0000-0001-7130-2880]{Leonardo \'Ubeda}
\stsci

\author[0000-0002-0012-2142]{Thomas G. Williams}
\oxford

%------------------------INTRODUCTION-------------------------

\section{Introduction}
\label{sec:Introduction}

In this paper we leverage new, high physical resolution, high sensitivity maps of H$\alpha$ emission from nearby galaxies to assess the coincidence between recent supernovae (SNe) and \ion{H}{2} regions. This measurement can help constrain both the nature of SN progenitors and the environments into which SNe explode.

Bright regions of H$\alpha$ emission in galaxies are indicative of \ion{H}{2} regions, where short-lived ($\lesssim 5$~Myr), massive ($>$10 $M_\odot$) stars ionize the gas through photons with energies higher than 13.6 eV. Therefore we would expect SNe in close proximity to \ion{H}{2} regions to likely originate from massive, short-lived progenitors. Such proximity studies have been particularly useful for understanding if and how the different subtypes of core-collapse SNe (CCSNe), Types II, Ib, Ic, correspond to different progenitor mass and age ranges \citep[e.g.,][]{Anderson2015}, something that has been challenging to infer from direct photometry/spectroscopic observations of the SNe \citep[e.g.,][]{Dessart2022}.

SN environment studies attempt to constrain the delay time and progenitor populations for SNe by measuring the correlation of the different SN types with various stellar populations and tracers of star formation \citep[see][for an excellent review]{Anderson2015b}. In a key early study, \citet{vanDyk1996} found $\sim70\%$ of CCSNe to be associated with \ion{H}{2} regions. More recently, \cite{AudcentRoss2020} compared the radial distributions of 80 SNe with R-band, UV, and $H\alpha$ emission in the SINGG/SUNGG galaxy surveys. They found SNe Ia correlated with R-band light, an indicator of the presence of low mass progenitors; SNe II were correlated with FUV emission, consistent with moderately massive progenitors; and stripped-enveloped SNe (SESNe) were found to be the most associated with $H\alpha$ emission, suggestive of SESNe originating from the most massive progenitor systems. Recent studies expanded beyond only imaging to also leverage IFU data \cite[e.g.,][]{Anderson2014,Galbany2017}. These studies consistently show that SNe Ia (white dwarf SNe) are the most weakly associated with tracers of star formation and instead associated with the older stellar population \citep[e.g.,][]{Pritchet2024}. SNe Ib/c (stripped envelope core collapse SNe) are more associated with tracers of very recent star formation than SNe II (other core collapse SNe) \citep{Crowther2013, Galbany2014}, but with subtle differences between SNe Ib and Ic \citep[][]{James2006,Anderson2012}. These results support a picture with an increasing progenitor mass sequence, SNe Ia $\rightarrow $ SNe II $\rightarrow$ SNe Ib $\rightarrow$ SNe Ic \citep[but note that binary evolution can have a complicating effect on this simple picture; e.g.,][]{Kuncarayakti2013,Zapartas2017}.

The coincidence (or not) of SNe and \ion{H}{2} regions also has significant implications for the mechanisms and impact of stellar feedback in various contexts. A variety of observational evidence supports that cold, molecular gas is often substantially cleared from a star-forming region \textit{before} H$\alpha$ fades \citep[e.g.,][]{Schruba2010,Kruijssen2019,Schinnerer2019,Chevance2020,Kim2022,Pan2022}. Measurements of resolved H$\alpha$ emission around clusters with SED modeling-based ages also suggest that the ionized gas may be cleared on few Myr timescales, even before ionizing photon production stops \citep[e.g.,][]{Hannon2022}. If SNe have an important role in this rapid gas clearing, then they must be present to some degree within \ion{H}{2} regions. Arguments based on stellar evolution timescales ascribed most of this gas clearing to ``pre-SN'' feedback: stellar winds, radiation pressure, and ionized gas pressure \citep[e.g.,][]{Lopez2014, McLeod2019, McLeod2020, McLeod2021, Barnes2020, Olivier2021, Barnes2022, Chevance2022}. Directly measuring the coincidence of SNe with \ion{H}{2} regions provides an independent, empirical test.

Conversely, SNe explosions in galaxy disks are important to supporting the overall gas disk, launching galactic winds, and stirring turbulence \citep[e.g.,][]{Elmegreen2004,Ostriker2011,Walch2015,Girichidis2016,Veilleux2020}. The environment where a SN explodes has a significant impact on its zone of influence, with explosions in denser environments exerting a more local influence both because the high density leads to a shorter cooling time and because the momentum injected by the supernova affects a smaller physical region at high density \citep[e.g.,][]{Walch2015,Gatto2017,Keller2020}. Away from the dense gas of star-forming regions, SN explosions are free to impact a larger area and exert this ``large-scale'' feedback \citep[e.g.,][]{Barnes2023}. Therefore demonstrating what fraction of these explosions in fact do occur away from dense gas represents an important avenue to quantitatively understand feedback in stellar disks. Recently both \citet{MaykerChen2023a} and \citet{Sarbadhicary2023} measured the cold gas content (CO and \ion{H}{1}) at the sites of recent SNe or likely near-future SNe in nearby galaxies. They found evidence for substantial populations of SNe away from CO emission and well-positioned to explode into low density regions. Even SNe close to star-forming regions of galaxies can still go off in the low density environments found in bubbles carved out by stellar populations and previous SNe \citep[]{Bagetakos2011, Pokhrel2020, Barnes2023, Egorov2023, Watkins2023a, Watkins2023b}.

Because SNe are rare, SN environment studies have often been forced to work with samples of distant objects with relatively coarse physical resolution. This can make it difficult to distinguish cases where a SN occurs \textit{within} an \ion{H}{2} region from cases in which the SN only occurs \textit{near} the \ion{H}{2} region. Studies at higher resolution are needed to directly measure the fraction of SNe that are actually occurring \textit{within} \ion{H}{2} regions and to place these explosions accurately within the multiphase interstellar medium (ISM).

In this work, we adopt a slightly different approach compared to these earlier studies. While they often characterize the sites of SN explosions after detecting them, we instead focus on identifying SNe and determining their locations within a uniquely well-studied set of nearby $\lesssim 20$~Mpc galaxies, the PHANGS surveys \citep[][]{Leroy2021,Emsellem2022,Lee2022}. Because PHANGS targets very nearby galaxies, even seeing-limited ground-based observations achieve physical resolution $\lesssim 100$~pc, up to ten times sharper than previous studies. This offers the prospect to better localize SNe relative to \ion{H}{2} regions, and from the rich supporting data, we can construct a variety of careful controls using data at other wavelengths. This high physical resolution and multi-wavelength coverage offers the prospect to ``zoom in'' on individual SN sites to directly see the likely area of influence of the SN and so understand its future impact. 

We took a first step towards this goal of characterizing the local sites of SNe in \citet{MaykerChen2023a}, where we studied the CO~(2--1) emission from SNe in the PHANGS--ALMA survey \citep[][]{Leroy2021}. In this paper we take the next logical step: comparing SNe to tracers of the ionized gas and young, massive stars using the PHANGS--MUSE survey \citep[][]{Emsellem2022} and including a first look based on the PHANGS--H$\alpha$ HST survey \citep[P.I.\ R.\ Chandar; Chandar, Barnes et al. in prep.,][]{Barnes2022}. PHANGS--MUSE, our core comparison data set, provided spectroscopic optical mapping at $\lesssim 1\arcsec$ resolution for $19$ galaxies. This includes high quality maps of H$\alpha$ and H$\beta$ \citep[e.g.,][]{Belfiore2022,Belfiore2023} that have been used to identify and characterize $\sim 20,000$ individual nebular regions \citep[][]{Groves2023}. We identify 32 SNe that have occurred within 10 targets of PHANGS--MUSE and characterize their explosion sites, exploring implications for SN progenitors and stellar feedback.

In \S \ref{sec:Methods} we describe our experimental design and data. In \S \ref{Maps} we summarize the data used in this work. In \S \ref{SNselection} we discuss how our SN sample is compiled. In \S \ref{Analysis} we discuss the control measurements that we make to help interpret our results. In \S \ref{sec:ResultsHa} we report on the  H$\alpha$ emission at our SN sites: in \S \ref{sec:HIIregions} we measure the fraction of SN sites coincident with an \ion{H}{2} region, in \S \ref{sec:types} we analyze our SN population by type, in \S \ref{sec:SSP} we compare our findings with expectations of simple stellar population (SSP) models, in \S \ref{sec:HIIregionsDist} we compare the distance to the nearest \ion{H}{2} region from our real SN sample to three model populations of SNe,  in \S \ref{sec:HaEmission} we compare the distributions of H$\alpha$ emission present at the sites of the real SNe to our model populations, and then in \S \ref{sec:Zooms} we zoom in to examine the local MUSE and HST H$\alpha$ emission from our SN sites. In \S \ref{sec:ResultOther} we also consider other properties of the ISM at the sites of our SN sample. In \S \ref{sec:HACO} we look at how the incidence of molecular gas relates to the presence of \ion{H}{2} regions; in \S \ref{sec:BD} we compare the extinction at each of our SN sites to that of the host galaxies overall; and in \S \ref{sec:Additional} we report additional diagnostics at each of our SN sites, including velocity dispersion, metallicity, and BPT classifications. Finally, in \S \ref{sec:Discussion} we summarize and discuss our results.

%------------------------METHODS-----------------------------

\section{Methods}
\label{sec:Methods}

\subsection{Data}
\label{Maps}

\begin{figure*}[]
\centering 
\includegraphics[width=0.95\linewidth]{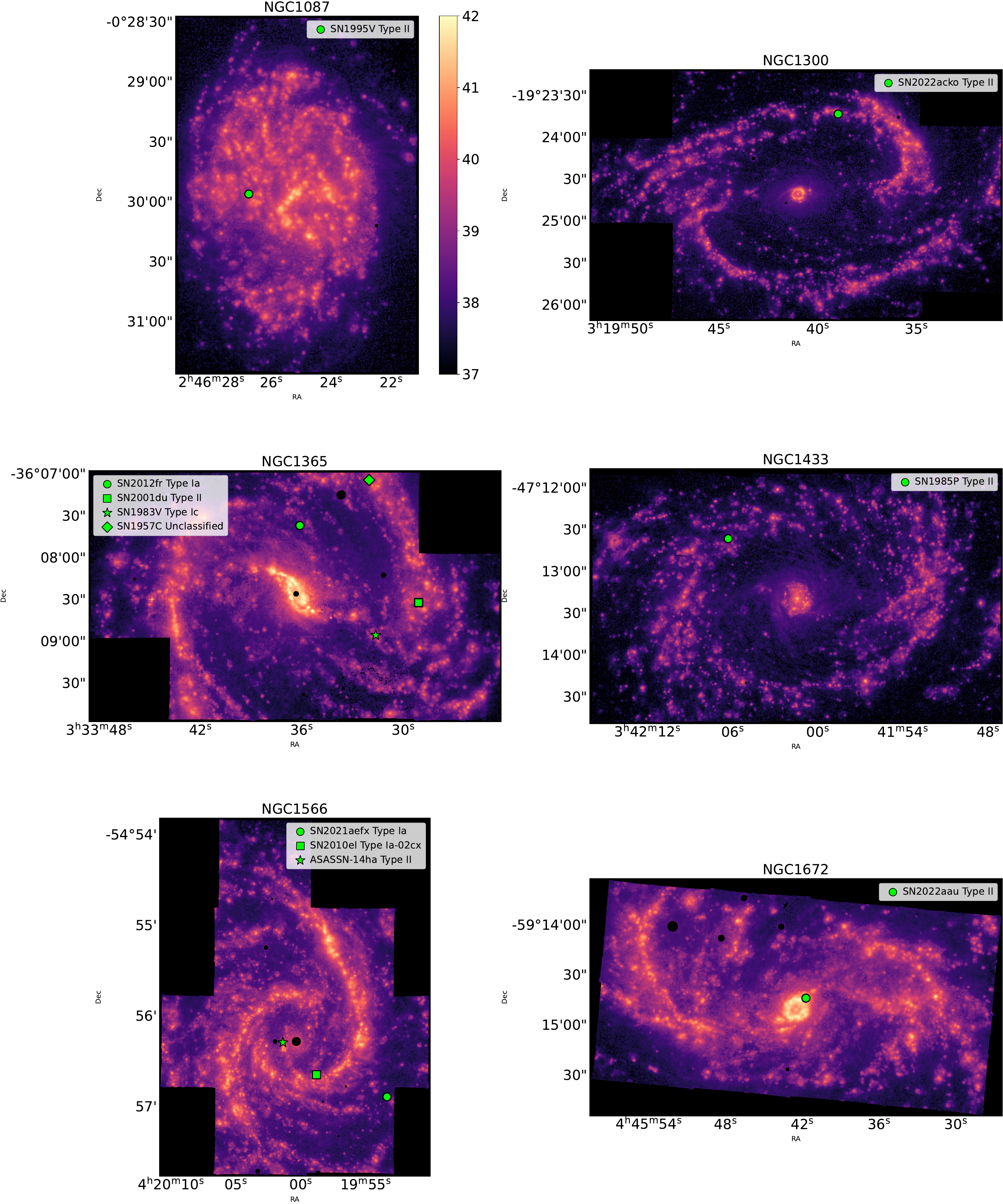}
\caption{\label{fig:MUSEGalaxies}
Recent SNe (green symbols) that occurred within the footprint of the PHANGS--MUSE survey \citep{Emsellem2022} plotted over PHANGS-MUSE images of extinction-corrected H$\alpha$ intensity on the same logarithmic stretch, shown in the colorbar. Black dots are where foreground stars have been removed from the image. Each SN in each galaxy is given a unique marker shape, with details of each galaxy given in Table~\ref{tab:galaxyTable} and each SN given in Table~\ref{tab:supernovaeTable}. The Appendix presents ``zoom in'' images around each SN and comparison to higher resolution HST H$\alpha$ mapping. Continued in Fig. \ref{fig:MUSEGalaxies_cont}.}
\end{figure*}

\begin{figure*}[]
\centering 
\includegraphics[width=0.95\linewidth]{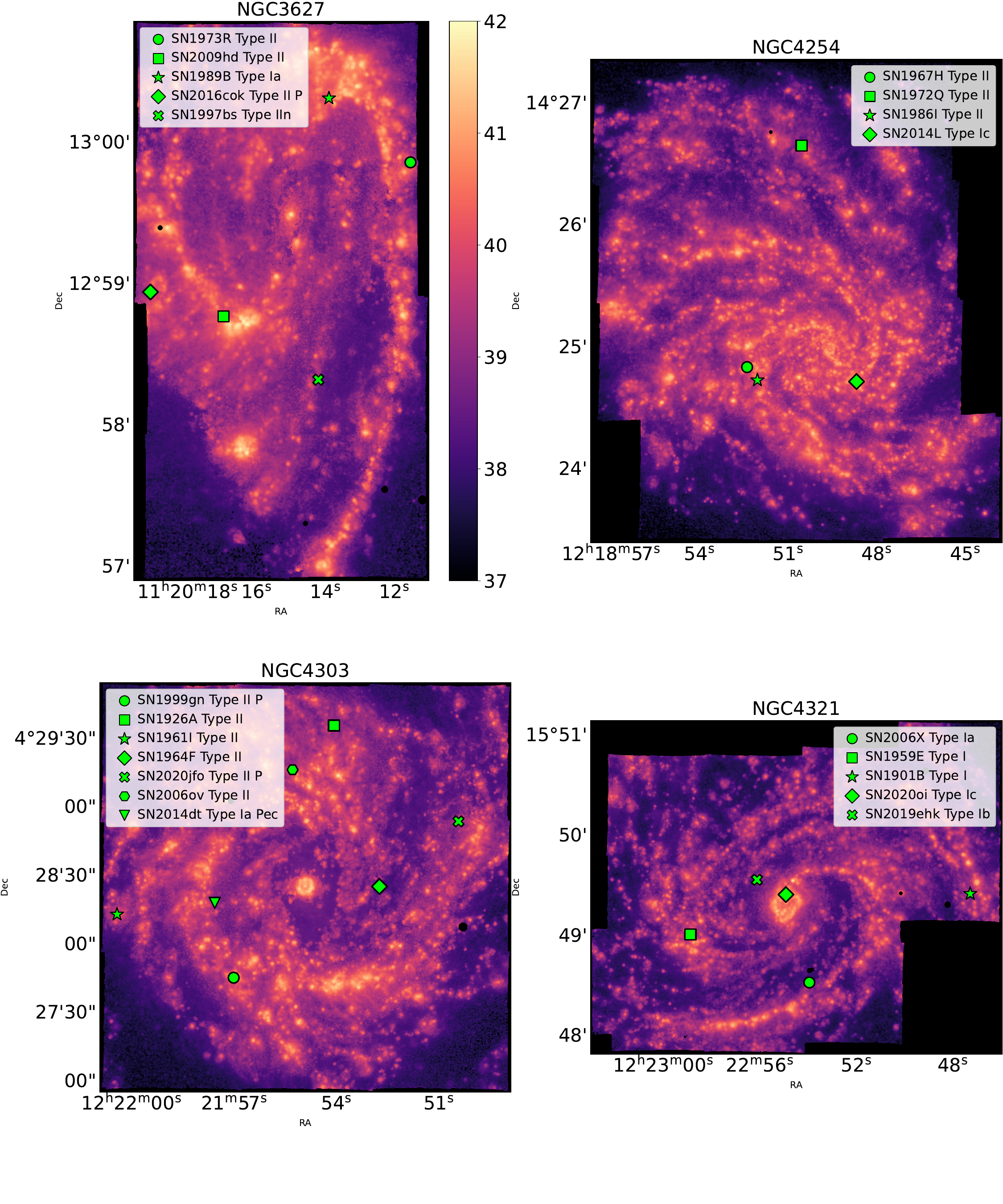}
\caption{\label{fig:MUSEGalaxies_cont} Figure~\ref{fig:MUSEGalaxies} continued.}
\end{figure*}

The PHANGS--MUSE survey \citep{Emsellem2022} mapped 19 nearby, star-forming galaxies (distances $<$ 20~Mpc, inclinations $<$ $60^\circ$) using the Multi-Unit Spectroscopic Explorer (MUSE) instrument \citep{Bacon2010} on the ESO Very Large Telescope. For this work, we use attenuation-corrected H$\alpha$ emission maps (see Figures \ref{fig:MUSEGalaxies} and \ref{fig:MUSEGalaxies_cont}) produced by \citet{Belfiore2023} \citep[see also][]{Pessa2021}. We use the ``copt'' convolved and optimized standard products. These have resolution that varies from galaxy to galaxy ranging between 0\farcs78 and 1\farcs25, which translates to linear resolution of $33{-}104$~pc at the distance of our targets (see Table~\ref{tab:supernovaeTable}). The MUSE astrometry is calibrated against Gaia via a procedure detailed in \citet{Emsellem2022} and so is expected to be accurate to within a few tenths of an arcsecond. 

The MUSE maps have high signal-to-noise, recovering H$\alpha$ emission almost everywhere. \cite{Emsellem2022} found that more than 95\% of  0\farcs2 spaxels within 0.5~$R_{25}$ contain H$\alpha$ emission at a 3$\sigma$ level. However, much of this emission represents diffuse ionized gas, likely powered by photons leaked from \ion{H}{2} regions \citep[e.g.,][]{Belfiore2022}. To distinguish likely actual \ion{H}{2} regions from this extended emission, we use the PHANGS nebular catalog \citep{Groves2023, Santoro2022} to determine where there is an \ion{H}{2} region along the line of sight. These \ion{H}{2} regions are identified using the \mbox{HIIphot} algorithm \citep{Thilker2000} adopting a single termination gradient of the H$\alpha$ surface brightness. For more information, see \S 3.1 of \cite{Groves2023}. 

In Table~\ref{tab:supernovaeTable}, we report the nebular catalog's BPT flags for three emission line diagnostics, \ion{[N}{2}] $\lambda$6584, \ion{[S}{2}] $\lambda$6717, \ion{[O}{1}] $\lambda$6300, \citep[for more information, see Section 4.2 of][]{Groves2023}. These diagnostics help determine whether the emission is more likely to arise from an \ion{H}{2} region or another type of nebula, e.g., supernova remnant shocks, winds, or planetary nebulae. As we discuss further in section \ref{sec:Additional}, all 13 SNe that are within the line of sight of nebular emission show diagnostics suggesting that the nebular emission is caused by star formation, and all have low H$\alpha$ velocity dispersions. Overall, the H$\alpha$ emission occurring within the line of sight to our SN sample is consistent with emission from \ion{H}{2} regions. Because our sample consists of young SNe, we do not expect that they have had time to influence their environment and the BPT diganostics are unlikely to be affected by the SNe themselves. The one exception is SN2017gax which was observed by PHANGS--MUSE shortly after the SN and contains the spectrum of the explosion. As a result, we remove this object from our sample.

In addition to the MUSE data, we compare SN locations to new narrowband H$\alpha$ emission maps from the \textit{Hubble Space Telescope} (P.I.\ R.\ Chandar; R.\ Chandar, A.\ Barnes et al. in preparation). These are similar to the map of NGC~1672 presented in \citet{Barnes2022} with similar processing applied. Because they are diffraction limited at $\approx 0\farcs1$ resolution (2.6-9~pc linear resolution), they offer a much sharper view of H$\alpha$ emission and \ion{H}{2} regions than VLT-MUSE but with worse surface brightness sensitivity. Because these data are quite new and still in a preliminary state, we use them here primarily for a qualitative comparison to the MUSE results in \S \ref{sec:Zooms}.

We trace molecular gas surface density using CO~(2--1) maps from the PHANGS--ALMA survey \citep{Leroy2021}. These data have similar resolution to the MUSE H$\alpha$ maps and we use them to assess the presence of cold, molecular gas in the vicinity of \ion{H}{2} regions where SNe are detected. CO~(2--1) emission is a standard tracer for the cold, dense, star-forming phase of the ISM \citep[for a review see][]{Bolatto2013}. Here we primarily focus on the detection of CO emission, reporting if there is significant (signal-to-noise $>$ 3) CO~(2--1) emission present. For a more detailed analysis of CO~(2--1) emission at the sites of SNe explosions in the PHANGS galaxies, with more description of how we handle the PHANGS--ALMA CO~(2--1) data, see \citet{MaykerChen2023a}.

We also use near-infrared (near-IR) emission to trace the overall distribution of stellar mass in our targets. This allows us to construct control measurements that predict the amount of chance coincidence between SNe and \ion{H}{2} regions expected for normal (not just massive) stars. We trace the surface density of stellar mass using near-IR ($3.6\mu$m) maps from the Infrared Array Camera (IRAC) on the \textit{Spitzer Space Telescope}. These were mostly obtained or reprocessed as part of the S$^{4}$G survey \cite{Sheth2010} and details of their processing and origin in the context of PHANGS is described in \citet{Querejeta2021}. We use the near-IR intensity for this purpose and do not make any corrections to account for local variations in the stellar mass-to-light ratio.

\subsection{SN selection}
\label{SNselection}

Following \citet{MaykerChen2023a}, we gather a population of recent ($<$ 125 years) SNe using the Open Supernova Catalog (OSC)\footnote{Accessed on 26 January, 2022, but no longer accessible online.} and the Transient Name Server (TNS)\footnote{https://www.wis-tns.org/ (accessed December 30, 2023)} . We are interested in SNe that have exploded recently and therefore have not had enough time to influence their surrounding environment. A typical early-stage SN shock with velocity of $10^4$~km~s$^{-1}$ \citep{Draine2011} will only have expanded $\sim$1 pc in 100 years, which is much smaller than the resolution of our data. The Open Supernova Catalog also records supernova remnants, which can confuse this selection. To avoid including these, we only select SNe that have a recorded discovery date. 

We select a population of 32 SNe within 10 galaxies in the PHANGS--MUSE galaxy footprints. Table~\ref{tab:galaxyTable} lists the SNe along with their host galaxy, type, right ascension and declination, whether they are included in our project sample, and the reference paper used for their type classification. A total of 36 SNe in 11 galaxies were originally identified, but we do not analyze two SNe (SN2013ej in NGC~0628 and SN1979C in NGC~4321) because they lie just outside of the map coverage. We remove SN2019krl in NGC~0628 because of uncertainty in its type classification and the likelihood that it is instead a non-terminal explosion \citep{Andrews2021}, and finally we remove SN2017gax in NGC~1672 because it was imaged shortly after explosion and the SN's light has dominated the spectrum. This reduces our working sample to 32 SNe within 10 galaxies, $19\%$ (6/32) are SNe Ia, $59\%$ 19/32 are SNe II, $13\%$ (4/32) are SESNe, and $9\%$ (3/32) are unclassified. We show their locations in Figure~\ref{fig:MUSEGalaxies}. In Table~\ref{tab:supernovaeTable} we report the type classification, native H$\alpha$ map resolution, H$\alpha$ intensity measured at the SN site, the velocity dispersion, extinction, \ion{H}{2} region status, BPT diagnostic line classifications, presence of CO~(2--1) emission, directly measured \ion{H}{2} region metallicities (when available) and calculated metallicities using galaxy gradients, galactocentric radius, and effective radius of each SN site. 

The measurement of the spatial coincidence of SNe and H$\alpha$ emission is limited by the resolution of the MUSE data and the positional accuracy of the SN location. The MUSE resolution usually represents the limiting factor for SNe detected in the last few decades, but for the older half of our sample the location of the SNe may also contribute uncertainty. 

Assessing positional uncertainties in the SNe precisely is complicated because the OSC, TNS, and the vast majority of SN discovery papers do not report positional uncertainties. To estimate a typical positional uncertainty for our SNe, we examine a wide selection of individual SN discovery papers and observe the change in the reporting confidence from discoveries that took place in the mid-1900s and those that have occurred in the last few years. We find that SNe that have occurred in the last $\sim20$ years are more likely to report uncertainties in their position---ranging from 0.1 to 1\arcsec\ \citep[e.g., see][]{Evans2003, Monard2008, Pignata2009} and recent SN discovery papers that do not report uncertainty, their position is reported to the nearest 0.1 to 1\arcsec\ as well. Because of this, we assume that for modern SN searches like ASAS-SN \citep{Shappee2014, Kochanek2017}, a typical positional uncertainty will be  $\lesssim 1$\arcsec. However positional uncertainties for early SNe, such as SN1926A, could be as high as $\sim10$\arcsec.

Half (16/32) of our sample has occurred within the last 20 years. We consider that for these, the positional uncertainty is of the same order as the angular resolution of MUSE maps, while for the remaining half, the uncertainty in the astrometry is dominated by the uncertainty in the reported SN location.

\subsection{Relative positions of SNe, HII regions, and statistical controls}
\label{Analysis}

We measure whether each SN occurs inside a known \ion{H}{2} region, note the distance to the nearest pixel in the \ion{H}{2} region\footnote{The borders of \ion{H}{2} regions correspond to \mbox{HIIphot} masks.} for SNe outside \ion{H}{2} regions, and record the intensity of H$\alpha$ emission at each SN. Because our galaxies have extended H$\alpha$ emission and often have a large fraction of their area covered by \ion{H}{2} regions, we also construct several control scenarios to assess the probability of random coincidence. Similar to \citet{MaykerChen2023a}, we use several model populations as controls to help interpret our measurements for our real SN sample. We consider these cases:

\begin{enumerate}
\item \textit{Purely random:} In this case, SNe are equally likely to occur in each pixel of the map, leading to a purely random distribution.

\item \textit{Random within the local region around the SN:} We also consider a scenario where the SN occurs randomly at a position within a $500~\textrm{pc} \times 500~\textrm{pc}$ or $1~\textrm{kpc} \times 1~\textrm{kpc}$ box centered on the SN explosion. This is intended to capture that some other variable might cause SNe to explode in a general region of the galaxy and to test for chance small-scale coincidence between \ion{H}{2} regions and SNe once the general SN location is set.

\item \textit{Following the H$\alpha$ distribution:} In this case, the likelihood for a SN to occur on any pixel is proportional to the intensity of H$\alpha$ emission at that location in the MUSE maps. We expect that our population of CCSNe to be better represented than the SNe Ia for this model due to the short delay-times of their progenitors.

\item \textit{Following the stellar disk traced by near-IR emission:} In this final control case, SNe are drawn based on the intensity of the near-IR maps described in \S\ref{sec:Methods}. This helps model SNe that trace the distribution of stellar masses, which should be appropriate for populations with longer delay times, e.g., SNe Ia \citep{Maoz2014, Anderson2015a, Cronin2021}.

\end{enumerate}

\begin{figure*}[ht!]
    \centering 
    \includegraphics[width=0.9\linewidth]{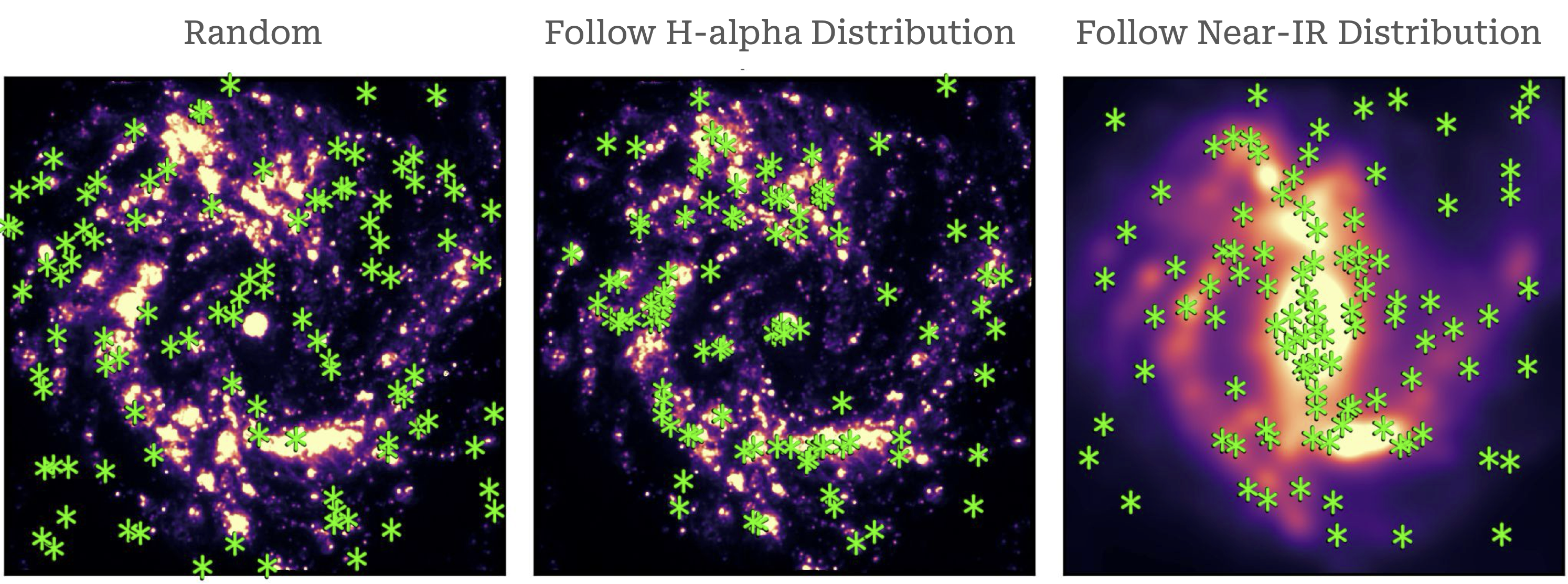}
\caption{Three models for SN placement in galaxies, each illustrated by placing 100 model-generated SNe in NGC~4303. From left to right - Model 1: the SNe are randomly placed across the footprint of the MUSE H$\alpha$ map with equal probability assigned to each pixel. Model 2: the  SNe are placed based on a probability distribution that follows the distribution of H$\alpha$ intensities in the MUSE H$\alpha$  map. Model 3: the SNe are placed based on a probability distribution that follows the near-IR light in the \textit{Spitzer} 3.6 $\micron$ map and therefore approximately traces the overall distribution of stellar mass.}
\label{fig:Models}
\end{figure*}

We generate 1,000 model SNe for each SN in our sample. This gives us a total of 32,000 model SNe from 10 galaxies generated for each model. We calculate the expected coincidence between these model distributions and the \citet{Groves2023} \ion{H}{2} regions. Following \citet{MaykerChen2023a} we also calculate the expected overlap with CO emission and with both CO emission and the presence of an \ion{H}{2} region in the line of sight for each case. Figure~\ref{fig:Models} shows examples of these control distributions for one galaxy. Table~\ref{tab:types} reports the coincidence between SNe and \ion{H}{2} regions expected for each of these models. We also compare the cumulative distributions of the models to our real SN sample in Figures \ref{fig:Distances} and \ref{fig:TypeIntensityCDFs}.

%------------------------RESULTS ----------------------------

\section{Results comparing H$\alpha$ emission to SN location}
\label{sec:ResultsHa}

In Figures \ref{fig:MUSEGalaxies} and \ref{fig:MUSEGalaxies_cont} we plot the PHANGS--MUSE H$\alpha$ maps for our sample of galaxies with SNe. We show the maps at their native resolution, which ranges from $0.78{-}1\farcs16$, and mark the locations of recent SNe in each galaxy. In this section we analyze the fraction of SNe that appear coincident with an \ion{H}{2} region (\S \ref{sec:HIIregions}) and compare these results with expectations from single stellar population (SSP) models (\S \ref{sec:SSP}), measure the distance of SNe from \ion{H}{2} regions (\S \ref{sec:HIIregionsDist}, \ref{sec:types}), the intensity of H$\alpha$ emission at the sites of SNe (\S \ref{sec:HaEmission}), and the relative position of H$\alpha$ emission and SNe at high resolution (\S \ref{sec:Zooms}).

%%%%%%%%%%%%% Section 3.1  %%%%%%%%%%%%%
\subsection{Fraction of SNe in \hiiforsec\ regions}
\label{sec:HIIregions}

\begin{figure}[ht!]         \includegraphics[width=0.9\linewidth]{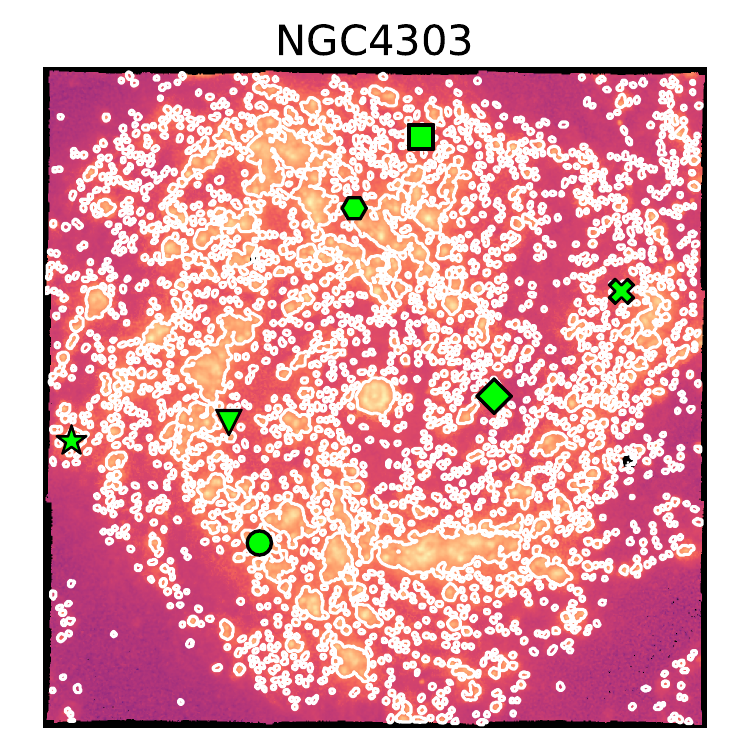}
\caption{\ion{H}{2} regions and recent SNe in NGC4303. As Figure~\ref{fig:MUSEGalaxies_cont} for NGC~4303 but now the white contours show the extent of the nebular region catalogs from \citet{Groves2023} and \citet{Santoro2022}.}
\label{fig:4303HII}
\end{figure}

As described in \S \ref{sec:Methods}, we use the \citet{Groves2023} nebular catalog to identify pixels where \ion{H}{2} regions lie along the line of sight through the galaxies. Figure~\ref{fig:4303HII} illustrates these \ion{H}{2} region contours for one target, NGC~4303. It shows that \ion{H}{2} regions cover a large area of the disk of the galaxy and encompass much of the bright H$\alpha$ emission. Across our sample, about $74\%$ of the total H$\alpha$ emission arises from pixels identified as \ion{H}{2} regions in the native resolution maps. The rest of the emission is associated with the extended diffuse ionized gas component \citep[e.g., see][]{Belfiore2022}. 

Table~\ref{tab:types} summarizes our basic results. We find that 41\% (13/32) of our SNe occur coincident with an \ion{H}{2} region. For comparison, across our galaxy sample, 13.7\% of map pixels are coincident with \ion{H}{2} regions. This means that if assigned to pixels randomly, 1/7 (4.4/32) of all SNe would occur along a line of sight that intersects an \ion{H}{2} region. 

The PHANGS--MUSE images often cover a large portion of a galaxy, including both the dense, high surface density inner regions and spiral arms and the more extended, lower surface density regions. As a result, some of the observed coincidence between SNe and \ion{H}{2} regions may result from the SNe simply occurring in denser parts of galaxies where both \ion{H}{2} regions and stars are more common. To account for this, we also construct a set of more localized comparisons, in which we compare the fraction of SN locations with detected \ion{H}{2} regions to only the fraction of area covered by \ion{H}{2} regions in nearby pixels. Here we define ``nearby'' as boxes $500 \times 500$~pc or $1 \times 1$~kpc in size centered on each SN. The goal with this more stringent control is to test the idea that SNe occur specifically concentrated within HII regions.

Adopting this more stringent control, the results indicate that 32\% of pixels within $1~{\rm kpc} \times 1~{\rm kpc}$ local regions centered on our SN sites belong to a \citet{Groves2023} \ion{H}{2} region. This fraction increases to 36\% if we instead consider fields of size $500~{\rm pc} \times 500~{\rm pc}$. This means that if we allow that SNe are occurring in a given part of the galaxy, then we expect that $\sim 32-36\%$ will occur along the line of sight to an \ion{H}{2} region simply due to random chance. 

As an alternative control, we also examine the fraction of total near-IR emission arising from within the \ion{H}{2} regions. The near-IR traces the overall distribution of stellar mass. We find that 32\% of near-IR emission emerges from regions coincident with \ion{H}{2} regions. Similar to the ``local region'' control, treating the near-IR as a control distribution suggests that we would expect a third of our SN sample, or about 10 SNe, to lie within the \ion{H}{2} regions just by coincidence alone.

Thus both the local region control or using the near-IR suggest that we might expect one-third of SNe to occur near an \ion{H}{2} region by chance. We emphasize that this calculation is focused on coincidence \textit{along the line of sight at the MUSE resolution}. The actual position of the SNe along the line of sight is uncertain and the MUSE sizes may represent overestimates of the true \ion{H}{2} region sizes \citep[e.g.,][and see below]{Barnes2022}. On the other hand, it is still of physical interest that so many SNe appear to be located \textit{near} (rather than in) \ion{H}{2} regions by coincidence. This proximity provides an opportunity for interactions between SNe and ionized gas interactions during the SN remnant phase. SN explosions in or near \ion{H}{2} regions, even if by coincidence, also represent locations where different modes of feedback potentially amplify one another, indicating places where multiple generations of SNe are able to clear out larger volumes of gas. 

Our measured SNe-\ion{H}{2} region coincidence of $\approx 41\%$ is higher than either the $\approx 32{-}36\%$ suggested by the local models or the $\approx 32\%$ suggested by the starlight distribution. This indicates that some SNe \textit{do} preferentially occur in \ion{H}{2} regions, but also that controlling for random coincidence is critical to estimate the rate. Using a binomial distribution we estimate the uncertainty due to stochasticity to be $\pm 8.7\%$. 

Contrasting these local covering fractions with the fraction of SNe coincident with \ion{H}{2} regions, we find a general excess of 7.6\% $\pm$ 8.7\% of all SNe to be associated with \ion{H}{2} regions. In the next section, we separate our SN sample by type, specifically focusing on CCSNe which we expect to be more associated with star forming regions of galaxies.

%%%%%%%%%%%%%%%.  3.2 %%%%%%%%%%%%%%%%%%%%%%%%%%%
\subsection{Breakdown by type}
\label{sec:types}

%%%%%%%%%%%%% \ion{H}{2} region table %%%%%%%%%%%%%%%%%%%%%%%%
\begin{table*} 
\large
\centering
\begin{tabular}{lcccc}
\hline
Type & \ion{H}{2} region & \ion{H}{2} $+$ CO & CO only & Excess \\ 
\hline\hline
All SNe (32) & 41\% (13) & 34\% (11) & 22\% (7) & 7.6\% $\pm$ 8.7\% \\  
\hline
SESNe: IIb/Ib/Ic (4) & 100\% (4) & 75\% (3) & 0\% (0) & 38\% $\pm$ 23.5\%$^{a}$\\
SNe II (19) & 42\% (8) & 37\% (7) & 21\% (4)  & 10.5\% $\pm$ 11.3\%\\
SNe Ia (6) & 17\% (1) & 17\% (1) & 17\% (1) \\
Unclassified (3) & 0\% (0) & 0\% (0) & 67\% (2) \\
\hline
H$\alpha$ Emission & 73\%  & 69\% & 16\% \\
Near-IR (3.6\micron) Emission & 32\% & 23\% & 28\% \\
Random Map Pixel & 14\% & 9\% & 22\%  \\
Random Local (1~kpc) & 32\% & 26\% & 34\% \\
Random Local (500~pc) & 36\% & 30\% & 31\%  \\
\label{tab:types}
\end{tabular}
\centering
\caption{Occurrence of SN sample (top half), emission and pixel counts (bottom half) in the line-of-sight of ISM \\
\footnotesize a) Probability of control selecting 4/4 sites in \ion{H}{2} regions is 1.2\%} 
\end{table*}

Table~\ref{tab:types} shows the fraction of SNe coincident with \ion{H}{2} regions broken down by type. We find that 20\% (1/6) of SNe Ia, $42\%$ (8/19) of SNe II, and 100\% (4/4) of our stripped envelope SNe (SESNe; Types II, Ib, \& Ic) occur coincident to \ion{H}{2} regions. The sense of these results are consistent with that found by previous lower resolution work \cite[e.g.][]{Anderson2012, Anderson2014, Anderson2015, AudcentRoss2020}, which identified that increasing progenitor mass leads to increasing association with bright H$\alpha$ emission. As a reminder the progenitor mass sequence from lowest to higher is believed to be SNe Ia $\rightarrow$ SNe II $\rightarrow$ SESNe .

The mild 6\% excess of SN in \ion{H}{2} regions becomes more prominent when we consider only CCSNe. CCSNe (II and SESNe) have an excess of 19.2\% $\pm$ 10.4\% coincident to an \ion{H}{2} region, and when considering only SESNe we find an excess of 38\% $\pm$ 23.5\%. Although the SESNe sample is extremely small (only 4 SNe) the probability of our controls randomly placing SNe coincident to an \ion{H}{2} region 4/4 times is only 1.2\%. The physical association is also expected, because SESNe are thought to originate from the most massive stars. These will also be the earliest stars to explode. They may therefore be within or near to their birth sites, and likely to still be bright in $H\alpha$ emission. 

Thus, with the caveat that our controls suggest that random overlap may explain some of our measured associations, the differences among types matches physical expectations. Because \ion{H}{2} regions mark where young, massive stars have formed very recently, the SNe with the highest mass progenitors should be more likely to occur near these regions as they explode first. \ion{H}{2} regions have lifetimes of $\lesssim 5{-}10$~Myr, similar to the lifetime of a single massive star. Lower mass CCSNe with delay times of $\sim 10{-}30$ Myr are more likely to outlive their birth \ion{H}{2} region and will have more time to migrate away from the high density regions where they formed. We do not expect SNe Ias to have association with star-forming regions due to their long delay times, but we do expect that a fraction of SNe Ias will occur coincident with an \ion{H}{2} region just because the \ion{H}{2} regions cover an appreciable fraction of the galaxy. 

We note that none of the unclassified SNe occur in \ion{H}{2} regions. By contrast, \cite{MaykerChen2023a} found that most of the unclassified SNe occurred in regions that did have bright CO emission, but they worked with a sample about twice as large as the current work. 

Finally, we note that our limited sample size might lead to a sample that does not accurately reflect the broad SN landscape. Of specific interest here, previous SN surveys find that SESNe make up 25-30\% of CCSNe \citep{Smartt2009, Li2011a}, while only 12.5\% of our sample are SESNe. Given the high degree of association that we observe between the SESN and the \ion{H}{2} regions here, a more representative sample with more SESN would likely show a somewhat higher overall coincidence between SNe and \ion{H}{2} regions.

%%%%%%%%%%%%%%%%%% 3.3 %%%%%%%%%%%%%%%%%%%%%%%%
\subsection{Comparisons to expectations from SSP Models}
\label{sec:SSP}

%%%%%%%%%%%%%%%%%%%%%% Figure %%%%%%%%%%%%%%%%%%%%%%%%%

\begin{figure*}[ht!]
    \centering 
    \includegraphics[width=0.95\linewidth]{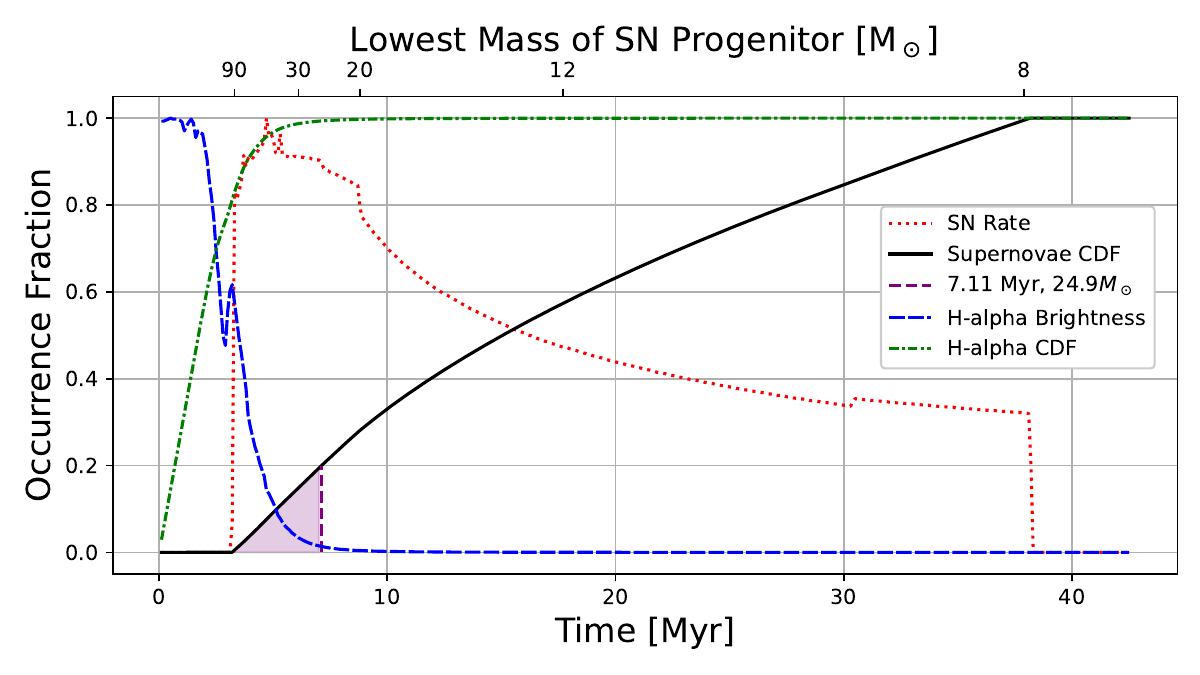}
\caption{Cumulative distribution function (CDF) of SNe over time from a simple stellar population modeled with STARBURST99. The CDF of the SNe is plotted with a solid black line, the normalized SN rate is plotted with a dotted red line, the CDF of H$\alpha$ emission is plotted with a dot-dashed green line, and the normalized H$\alpha$ brightness with a dashed blue line. In purple, we shade the SN CDF to indicate when it reaches the level of excess CCSNe (20$\%$) that we find coincident to \ion{H}{2} regions. We also mark the delay-time (7.11 Myr, 24.9$M_\odot$) that account for that excess with a dashed purple vertical line.}
\label{fig:SNtime}
\end{figure*}

Our analysis suggests that $\sim20\%$ of CCSNe preferentially explode in \ion{H}{2} regions. How does this compare to expectations? In Figure~\ref{fig:SNtime} we plot results for a simple stellar population (SSP) using the default assumptions from STARBURST99 \citep[SB99][]{Leitherer1999,Leitherer2014}. We plot the cumulative distribution functions (CDFs) as a function of time for both number of SNe explosions  and H ionizing photon production, which should map to H$\alpha$ emission.

In the SB99 model, the delay-time for CCSNe ranges from $3-37$~Myr, with the first $20\%$ of SNe occurring by 7.11~Myr. This is consistent with the H$\alpha$ brightness drop off which occurrs significantly by $\approx$7~Myr and the CDF of H$\alpha$ emission (or H ionizing photon production) shows that most of the H$\alpha$ has been produced before that time. 

If star-forming regions in our targets are indeed accurately described by an ensemble of SSPs, then our inferred SNe associated with H$\alpha$ would likely need to be associated with high-mass progenitor stars with lifetimes shorter than $\approx$7~Myr or with a binary core collapse SN production channel that operates on a similar timescale. To be precise, the first $20\%$ of the SNe in this SSP calculation occur by $7.11$~Myr and represent stars with masses $>24.9$~M$_\odot$.

In reality, the picture will be more complex than this. Supernova explodability work by \cite{Sukhbold2016} shows that the most massive stars have a much lower frequency of exploding, although there is no clear-cut boundary between explosions and non-explosions, the probability of successful explosions is less as you go above $20~M_\odot$, with the most massive stars requiring significant mass loss in order to explode. On the other hand, binary star systems extend the H$\alpha$ emitting lifetime and open more complex mappings between progenitor mass and explodability and different evolutionary timescales \citep[e.g.,][]{Eldridge2011, Vartanyan2021,Nguyen2022,Patton2022}. Likely, many of the regions we examine are also not SSPs, but host multi-generation or extended-duration star formation \citep{Rahner2017}. Another caveat is that the Starburst99 models assume a fully-sampled IMF, which is likely valid for the galaxy as a whole but may not be true for many of the individual regions, depending on the mass of the powering star cluster.

%%%%%%%%%%%%% Section 3.4  %%%%%%%%%%%%%
\subsection{Distance from SNe to \hiiforsec\ regions}
\label{sec:HIIregionsDist}

\begin{figure*}%
    \centering
    \subfloat{{\includegraphics[width=0.375\linewidth]{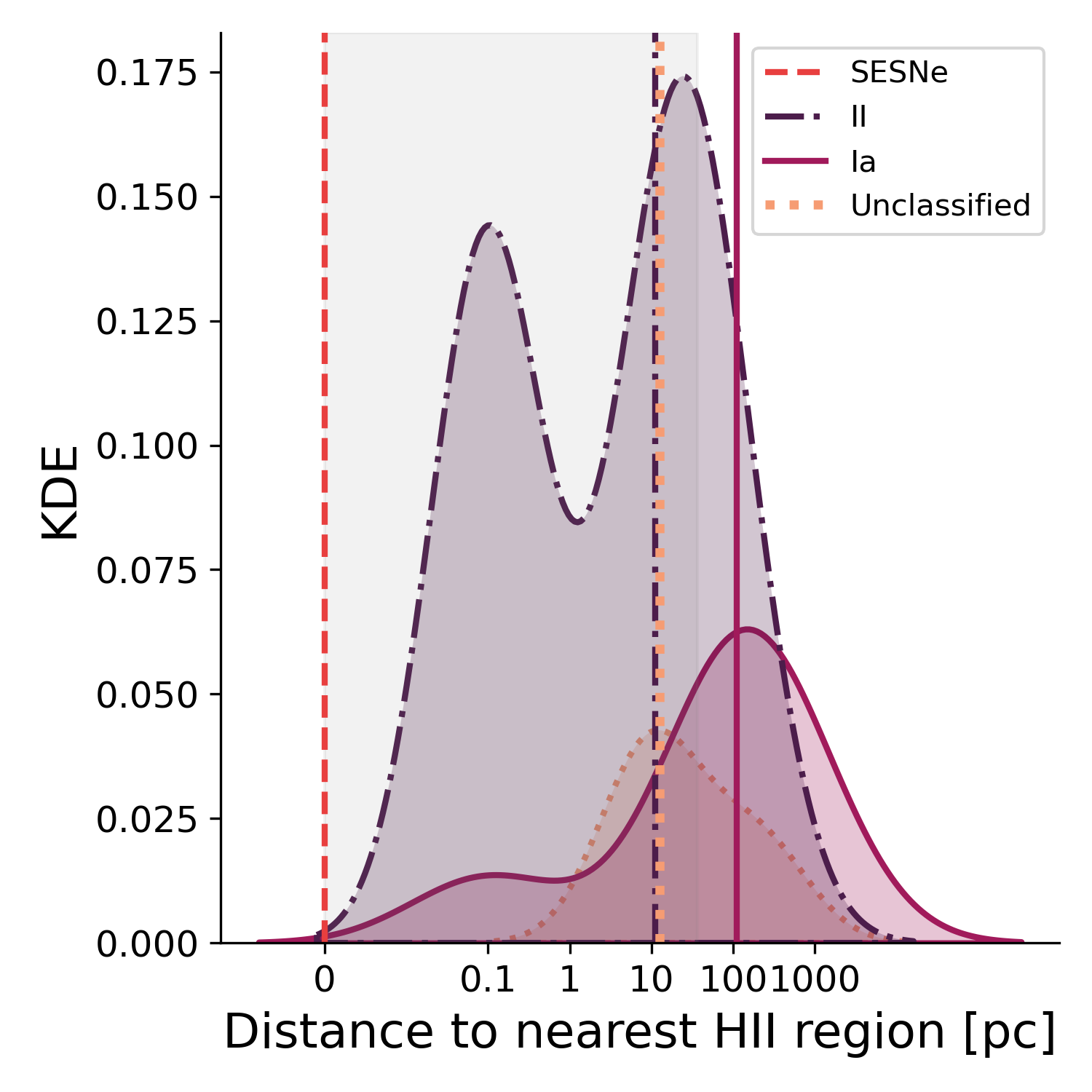} }}%
    \qquad
    \subfloat{{\includegraphics[width=0.525\linewidth]{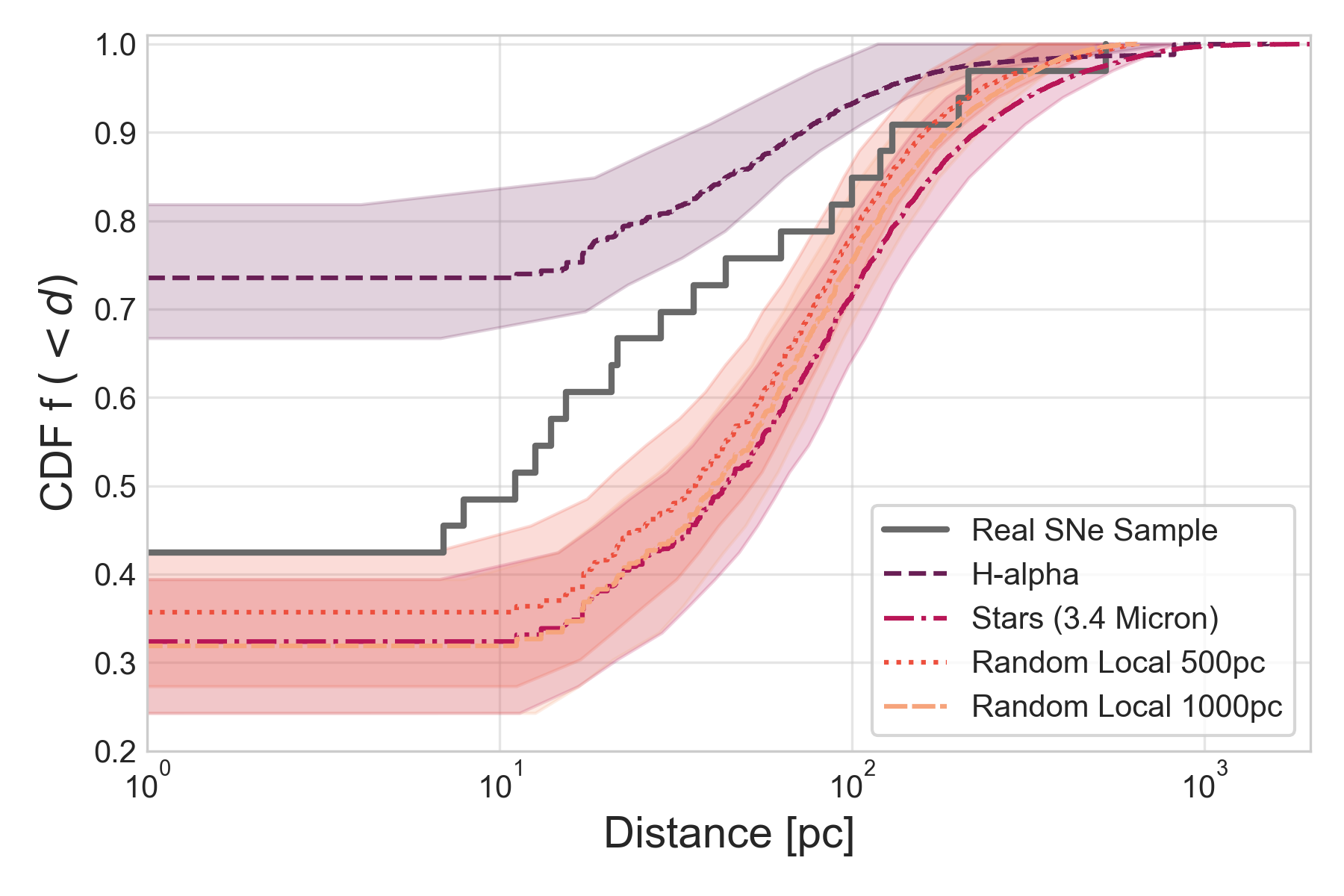} }}%
    \caption{\textit{Distances to nearest \ion{H}{2} region.} Left: kernel density estimation (KDE) of the distances from each SN site to the nearest \ion{H}{2} region. SNe with distances less than 1~pc are found within an \ion{H}{2} region. SNe II are represented with dash-dotted, dark-purple; SNe Ia with solid red-purple; SESNe with dashed magenta; and unclassifed SNe with dotted peach. Vertical color-and-style-coded lines mark the median distance value for each SN type. Grey shaded region marks the pixel scale. Right: CDFs of the distances to nearest \ion{H}{2} region for our real SN sample and three modeled populations. Our real SNe sample is drawn with a dark-grey line. The randomly generated SNe sample is drawn from the local (500 \& 1000~pc) map around each SN site. The 500~pc random pull sample is plotted with a dotted magenta line and the 1000~pc with a dashed peach line. The population generated from the H$\alpha$ distribution is plotted as a dashed, dark-purple line and the stellar disk distribution as dash-dotted red-purple line. The transparent shading represents the 16$^{\rm th}$--84$^{\rm th}$ percentile values from 1000 random pulls each the size of our observed SN sample from each model distribution. We find that our SESNe are overwhelmingly located in the line-of-sight to our \ion{H}{2} regions, while our SNe Ia sample tends to be less associated with \ion{H}{2} regions, and our SNe II are somewhere in between. Our real SN sample is distributed in between our H$\alpha$ model sample and our stellar disk and random local models.}%
    \label{fig:Distances}%
\end{figure*}

Even when SNe are not directly associated with \ion{H}{2} regions, we might expect those associated with very massive stars to be near such regions. To test this, we also calculate the distances to the nearest pixel tagged as an \ion{H}{2} region in the \cite{Groves2023} catalog for each of our SNe. 

The left panel of Figure~\ref{fig:Distances} shows the kernel density estimation (KDE) of the distances from each SN to the nearest \ion{H}{2} region, where the SN population is separated by type. Vertical color-and-style-coded lines mark the median distance value for each SN type. We shade the typical MUSE spaxel size in gray. The figure shows that our SESNe are always occurring coincident to \ion{H}{2} regions, our SNe II are more frequently found away from \ion{H}{2} regions than our SESNe, with a median distance of 10~pc. Although, our SNe Ia are often found farther away from \ion{H}{2} regions with a median distance of 110~pc, they are still relatively close. The overall low distances even for SNe not directly associated with \ion{H}{2} regions reflects that PHANGS-MUSE targets the actively star-forming parts of galaxies. As a result, H$\alpha$ emission is prevalent throughout the maps and even a SN associated with an older progenitor is still likely to occur close to an \ion{H}{2} region. 

Similar to \cite{MaykerChen2023a}, we compare the measured SNe-\ion{H}{2} region separations for our real SN sample to those produced for the model populations described in \S \ref{Analysis}. In the right panel of Figure~\ref{fig:Distances} we show the cumulative distributions of the distances to the nearest \ion{H}{2} region for both our real SN sample (32 SNe) and each of our four model SN populations (32,000 SNe generated for each model). We find that the distribution of our real SN population lies in between the model population that follows the distribution of H$\alpha$ emission and the model populations drawn from either the distribution of stellar mass or local regions around the real SN location. This reinforces the results above that while SNe are more associated with \ion{H}{2} regions than the general stellar population, they do not directly trace the distribution of H$\alpha$ emission. 

\subsection{Intensity of H\alphaforsec\ emission at the locations of recent SNe}
\label{sec:HaEmission}

\begin{figure}[ht!]
    \centering 
    \includegraphics[width=0.95\linewidth]{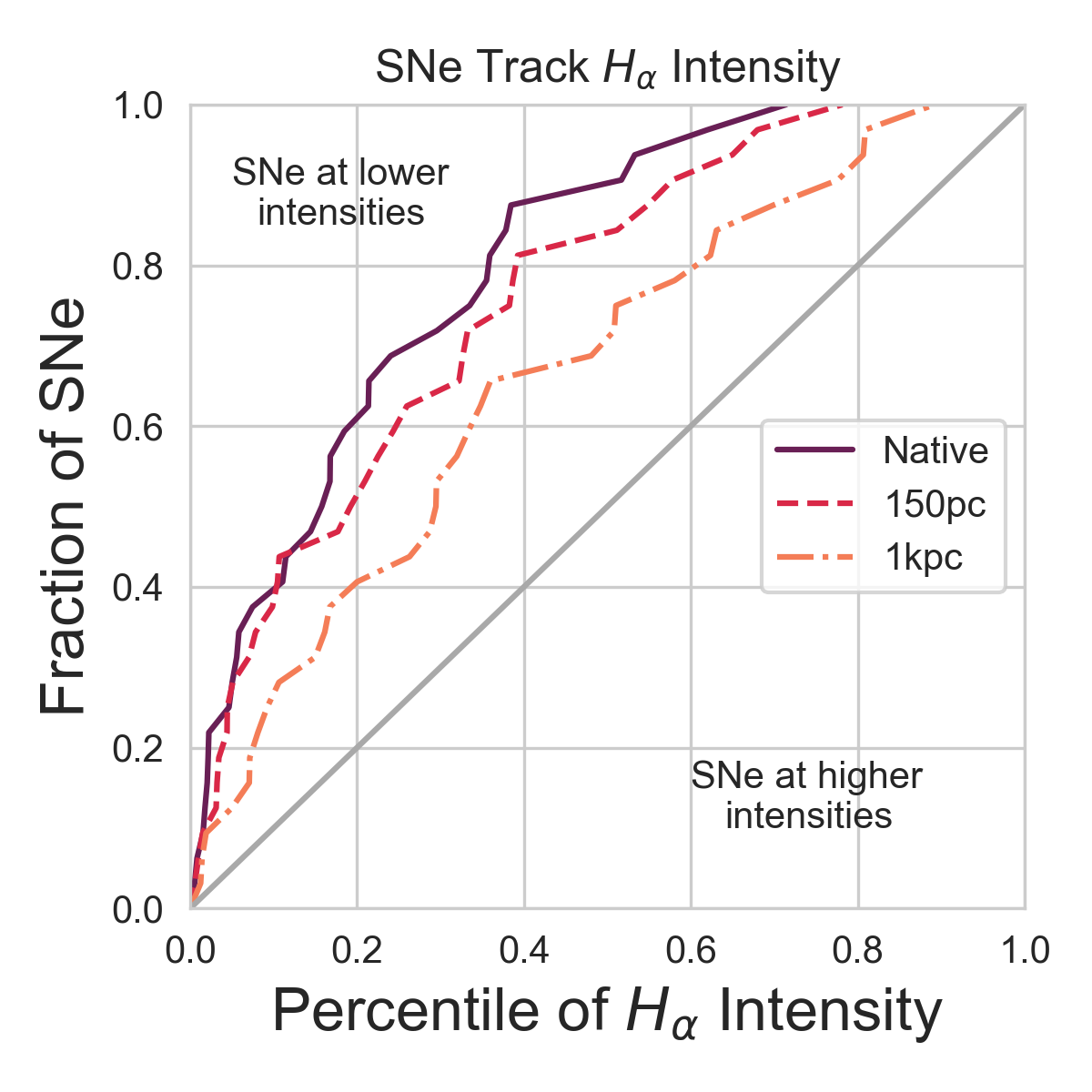}
\caption{Normalized cumulative rank (NCR) plot of the association of SNe with H$\alpha$ emission at native, 150pc, and 1kpc resolutions. Results for the PHANGS-MUSE native resolution (44--109~pc) are plotted with a solid purple line, 150~pc resolution results are plotted with a dashed magenta line, 1~kpc resolution results are plotted with a dash-dotted salmon line.}
\label{fig:resNCR}
\end{figure}

\begin{figure}[ht!]
    \centering     \includegraphics[width=0.95\linewidth]{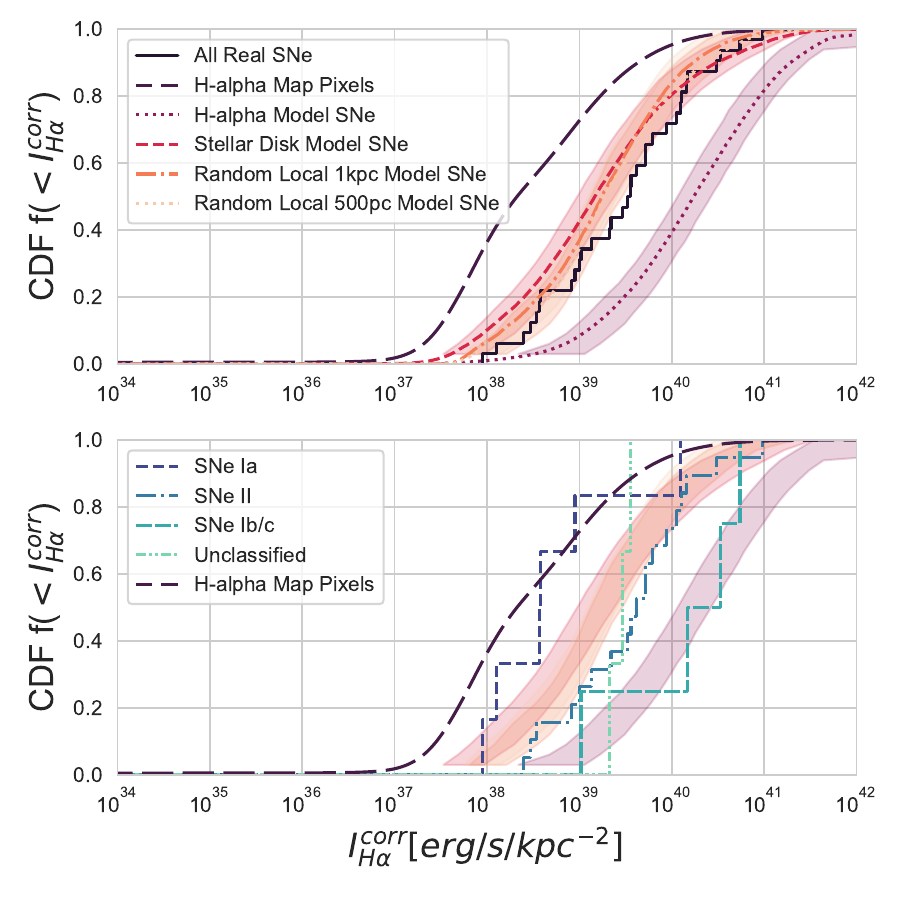}
\caption{Cumulative Distribution Functions (CDFs) of the extinction-corrected H$\alpha$ emission intensity distributions for our real SN sample, for all the map pixels and for our populations of model SNe. In the top panel, we compare the models to our whole real SN sample. In the bottom panel, we maintain the same comparison, but instead sort our SNe by type. The whole real SN sample (32 SNe) is drawn with a solid black line, the map pixels are drawn with a long-dashed, dark purple line. The H$\alpha$ model SNe are drawn with a dotted purple line, the stellar disk model SNe with a dashed pink line, and the random local models are drawn with orange dot-dashed and yellow dotted for 1kpc and 500pc local boxes respectively. Behind each model line we use transparent shading to represent the 16th--84th percentile values from 1000 random pulls, each the size of our observed SN sample, from each model distribution. The shaded percentiles are drawn on the lower plot as well but we remove the lines for readability. In the lower plot we sort our real SN by type, plotting the SNe Ia with a dashed dark blue lihe, the SNe II with a dot-dashed blue line, the SESNe (SNe Ib/c) with a closely dashed teal line, and the unclassified SNe with a dot-dot-dashed light green line. Note the increasing association with higher intensity H$\alpha$ emission as progenitor mass increases.}
\label{fig:TypeIntensityCDFs}
\end{figure}

So far, we have considered only whether a sight line is coincident with an \ion{H}{2} region. The intensity of H$\alpha$ emission provides additional information, tracking the production rate of ionizing photons and potentially conveying additional information on the distribution of massive, young stars within an individual \ion{H}{2} region. To leverage this information, we also measure the intensity of H$\alpha$ emission at the site of each of our SNe and report the results in Table~\ref{tab:supernovaeTable}.

Figure~\ref{fig:resNCR} shows the normalized cumulative rank (NCR) of SNe relative to the distribution of H$\alpha$ emission at native, 150~pc, and 1~kpc resolutions. An excellent overview of the NCR method is given in \citet{James2006}. Briefly, we sort the pixels in each H$\alpha$ map by H$\alpha$ intensity, assigning a percentile in the CDF to each. Then we note the percentile of the H$\alpha$ intensity CDF for the pixel at which each SN occurs and so construct the NCR of SNe relative to H$\alpha$. If the SNe occur in locations that track the distribution of H$\alpha$ intensity, then the measured curve will follow a line with slope unity. This analysis emphasizes coincidence between SN sites and \ion{H}{2} regions and could be impacted if the SNe and stellar winds cleared the surroundings of ionized gas. Although we do not expect our SN sample to have had time to dramatically influence their local environments, we address this in \S~\ref{sec:Zooms} where we plot the local ($500~\times~500~pc$) environments of each SN and look for signs of shell-like morphology in the H$\alpha$ emission.  

Consistent with the finding above that SNe appear in H$\alpha$-rich parts of galaxies, but not necessarily within \ion{H}{2} regions themselves, Figure~\ref{fig:resNCR} shows that our SNe sample tracks the H$\alpha$ emission more directly as the resolution becomes coarser. The fact that the lines are above and to the left of the one-to-one line means that actual SNe occur at lower $I_{H\alpha}$ on average than we would expect if the probability to find a SN tracked the distribution of H$\alpha$ emission exactly. The better agreement at coarser resolution reflects that at lower resolution, the beam of the telescope captures more of a region-average measurement of star formation activity, while at higher resolution we are able to better isolate individual \ion{H}{2} regions, and we observe that SNe appear somewhat offset from these regions. 

In Figure~\ref{fig:TypeIntensityCDFs} we plot the CDF of H$\alpha$ intensities at the sites of our real SN sample (32 SNe), the CDF for all H$\alpha$ emission at the native (44--109~pc) resolution of the MUSE maps, and the CDF of H$\alpha$ intensity generated for each of the the three model populations described in \S \ref{sec:Methods}. In the top panel we show results for our whole SN sample, and in the bottom panel we separate the SNe by type. Shaded regions represent the $16^{\rm th}{-}84^{\rm th}$ percentile range covered by repeat realization of the models. 

Similar to results with the distances from \ion{H}{2} regions, we find that the distribution of H$\alpha$ intensities at the real SN sites falls somewhere between the distributions for the model SNe generated from the near-IR light and the model SNe generated from the H$\alpha$ intensity. We expect that the higher mass progenitors would have a closer association with the H$\alpha$ distribution, while the lower mass CCSNe and SNe Ia would have a distribution that more closely resembles that taken from the overall stellar population traced by near-IR emission.

When we separate the SNe by type, we find the SNe Ia to exhibit a similar distribution to the map pixels or the near-IR emission. This agrees well with the expected lower mass range and wide age range of SN Ia progenitors. As the progenitor mass increases, we find an increasing association with higher H$\alpha$ intensity values, reaffirming the results of previous works \cite[e.g.][]{Anderson2012, Anderson2014, Anderson2015}. Of particular note, the SESNe (SNe Ib/c) show a distribution very similar to the H$\alpha$ maps themselves, supporting a direct association with high mass progenitors. 

\subsection{Zooming in on individual regions and comparison to \em{Hubble} Space Telescope imaging}
\label{sec:Zooms}

\begin{figure*}
\centering
\includegraphics[width=1.0\textwidth]{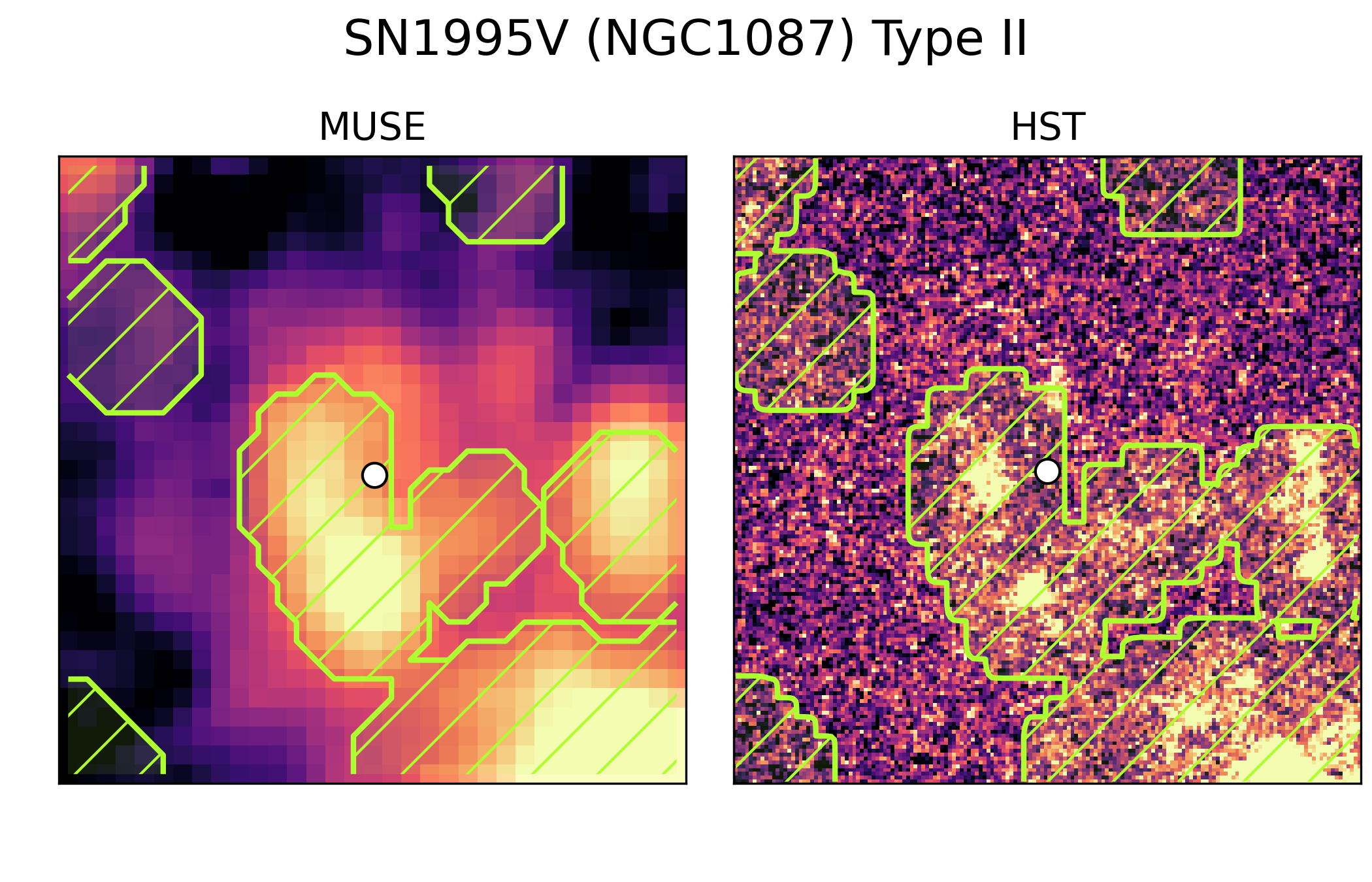} \caption{Example $500~\textrm{pc}\times500~\textrm{pc}$ zoom in for SN1995V (Type II) in NGC1087. The left panel shows H$\alpha$ intensity from the PHANGS-MUSE maps. The right panel shows H$\alpha$ emission from the same region at higher $\approx$0\farcs1 resolution from new narrowband HST mapping (Chandar, Barnes et al. in prep.) In both panels, the hatched green regions indicate the footprint of the \citet{Groves2023} nebular region catalog.}\label{fig:example_zoom}
\end{figure*}

In the figures in Appendix \ref{appendix}, we zoom in to each SN site, plotting $500~{\rm pc} \times 500~{\rm pc}$ cutouts of H$\alpha$ emission around each of the 32 SNe in our sample as well as marking the footprint of regions in the \citet{Groves2023} nebular catalog. We show an example of one such cutout in the left panel of Figure~\ref{fig:example_zoom}. These cutouts demonstrate that many of these SNe appear in or near regions rich in H$\alpha$ emission. They also show that, while many SNe occur within the footprint of the nebular regions, none are actually coincident with the local peak of H$\alpha$ emission. Instead, the SNe tend to lie off to the side, often near the edges of the regions, and are associated with lower intensity H$\alpha$ emission. 

To test this hypothesis further, we take advantage of very new, high-resolution ($\sim$0\farcs1) narrowband H$\alpha$ imaging of these targets using the \textit{Hubble} Space Telescope (P.I.\ R.\ Chandar; Chandar, Barnes et al., in preparation). As illustrated in the right panel of Fig. \ref{fig:example_zoom}, these data have been processed following a scheme similar to that described in \citet{Lee2022} and \citet{Barnes2022}. Here we use the data to provide a $\sim 10\times$ sharper view of the \ion{H}{2} regions compared to MUSE. Though the surface brightness sensitivity of HST does not match MUSE, the resolution of these data ($\sim 5{-}15$~pc) more closely matches the physical sizes of \ion{H}{2} regions seen in the Milky Way \citep[e.g.,][]{Anderson2014}.

In Figure~\ref{fig:HST} we remake the cutouts above using these HST H$\alpha$ data, again marking the locations of the \cite{Groves2023} \ion{H}{2} regions and significant CO~(2--1) emission. At the diffraction-limited, high resolution of HST, the \ion{H}{2} regions seen by MUSE are significantly more resolved, appearing smaller and significantly  better-defined. Many SNe located within the edges of \ion{H}{2} regions in the MUSE maps (e.g., SN1995V and SN1983V) appear offset from the \ion{H}{2} regions in the sharper HST maps. In fact, \textit{none} of the SNe with HST H$\alpha$ coverage actually lie on a high resolution H$\alpha$ peak, though we caution that the astrometric precision with which the SNe are located makes it hard to draw a firm conclusion.

Examining these images, especially the HST images, reinforces the statistical finding above that a large fraction of the SNe coincident with the \citet{Groves2023} \ion{H}{2} regions are likely to represent chance alignments. The overall active star-forming nature of the regions where the SNe occur and the relatively coarse physical resolution of the MUSE data lead to a large fraction of chance overlaps.

We might expect that pre-SN feedback would result in a shell-like morphology for the H$\alpha$, with lower densities of ionized gas surrounding the stellar population and near the site of the SN itself. However, such a morphology is not immediately clear from our images. We overwhelmingly see that our SN sites are off to the sides of the \ion{H}{2} region, rather than enclosed at least partially by shells. It is possible that the shells are too small to show up at the MUSE resolution and the lower sensitivity of the HST data. 

We caution that the positional uncertainty for the SNe represents a major limiting factor when comparing to the HST H$\alpha$ data. As Figure~\ref{fig:HST} shows, even in nearby galaxies, the angular size of individual \ion{H}{2} regions and nebulae is very small, so that placing the SNe precisely relative to these features requires astrometric precision of order 0\farcs1. Given this, future work that leverages HST and JWST to constrain the origin and impact of SNe will require high-quality astrometric positions. With this in mind, we are reassured that many of the SNe that show this characteristic near-but-not-in appearance have been recently discovered and have correspondingly secure astrometry. Our results suggest that SN feedback is not occurring at the centers of star-formation sites, and could have implications for clustered feedback simulations.

Finally, we note that in our previous work \cite{MaykerChen2023a}, we found a similar result when comparing the SN sites to molecular gas emission. The SNe associated with CO emission often appeared displaced from the CO peaks and near the edges of the detected regions. Those ALMA CO data have resolution $\approx 1\arcsec$, similar to the MUSE data. It will be telling to see whether at higher resolution the relative location of the SNe and the CO-traced molecular clouds similarly separate.

%%%%%%%%%%%%% Section 4  %%%%%%%%%%%%%

\section{Results comparing to other properties at the SN location}
\label{sec:ResultOther}

In \S \ref{sec:ResultsHa} we examined the coincidence of H$\alpha$ emission and recent SNe. We found a large fraction of SNe to occur along lines of sight that overlap \ion{H}{2} regions, but also showed that much of this overlap is likely to be coincidental. Both higher resolution imaging and statistical analysis suggest that many SNe occur near but not necessarily within \ion{H}{2} regions, with our best estimate that there is a $\approx 20\%$ excess of CCSNe associated with \ion{H}{2} regions relative to the controls.

In this section, we expand our analysis to consider other properties near the SN sites, examining the coincidence of H$\alpha$ and CO emission at SN sites (\S \ref{sec:HACO}), the extinction towards the ionized gas near SN sites (\S \ref{sec:BD}), and a variety of other diagnostics available from the \citet{Groves2023} and \citet{Emsellem2022} analysis of the MUSE data (\S \ref{sec:Additional}).

%%%%%%%%%%%%%%%%%%%%%%%%%%   SECTION 3.4  %%%%%%%%%%%%%%%%%%%%%%%%%%
\subsection{H\alphaforsec\ and CO}
\label{sec:HACO}

Newly-formed stars can remain embedded within their parent molecular clouds, with the mass of molecular material commonly traced by CO emission. In recent models of molecular cloud evolution, star-forming regions go through early phases in which newly formed stars still lie within their parent molecular cloud with H$\alpha$ emission almost undetectable \citep[e.g.,][]{Lockman1989, Kim2021}. Over time, the cloud is exposed to various forms of feedback which begin to disperse the natal cloud. This dispersal leads to phases in which the cloud is partially dispersed and H$\alpha$ and CO might be detected together and then as the cold gas becomes mostly dispersed, CO emission vanishes, but the H$\alpha$ emission remains fully visible \citep[e.g.,][]{Kawamura2009,Kruijssen2014,Kruijssen2018, Kim2022}.

By combining the MUSE H$\alpha$ maps with CO~(2--1) measurements from PHANGS--ALMA (see \S \ref{sec:Methods}), we identify which SNe in our sample occur coincident with only CO~(2--1) emission, an \ion{H}{2} region, both, or neither. To do so, we repeat our control scenarios listed in \S \ref{Analysis} and list our results in Table~\ref{tab:types}. We also overplot the locations of CO~(2--1) emission with a signal of 3 times the noise or higher in the zoom panels in Appendix~\ref{appendix}. Throughout, to count as a detection of CO~(2--1) emission, we require a signal-to-noise ratio of 3 in the integrated intensity maps masked with a ``broad'' mask\footnote{We are only considering CO~(2-1) detections in this work and not accounting for $\alpha_{CO}$ variations. See \cite{MaykerChen2023a} for a more detailed discussion of CO~(2-1) and SNe.}. Note that here we focus on the joint detection statistics for CO and \ion{H}{2} regions in this MUSE sample; \citet{MaykerChen2023a} show a more extensive analysis considering only CO for the full PHANGS--ALMA sample.

We find that $34\%$ of all SN sites are coincident with both an \ion{H}{2} region and CO~(2--1) emission, while $22\%$ are not associated with an \ion{H}{2} region yet are coincident with significant CO~(2--1) emission. Contrasting these with the control calculations, the CO-only sightlines appear largely consistent with random coincidence; that is, the percentage of real CO-only sightlines resembles that in the local or near-IR controls. Both the overall \ion{H}{2} region and the \ion{H}{2}+CO detection rates show an excess relative to these controls.

Similar to the \ion{H}{2} regions without CO~(2--1) emission, there is an increasing association of progenitor mass with \ion{H}{2} + CO~(2--1), with $17\%$ SNe Ia, $37\%$ SNe II, and $75\%$ SESNe found in \ion{H}{2} + CO~(2--1). However, this association breaks down when we consider only CO~(2--1) emission without a corresponding \ion{H}{2} region, which occurs for $17\%$ of SNe Ia, $21\%$ of SNe II, and $0\%$ of SESNe. Of note, $67\%$ of our unclassified SNe are found in CO-only regions, while none appear associated with \ion{H}{2} regions. This follows \cite{MaykerChen2023a} where the unclassified sample had a high correlation with dense CO~(2--1) gas. It is possible that these unclassified SNe are exclusively going off in embedded star-forming regions, and the occurrence in high extinction areas might account for the difficulty in providing a classification for the SN, but it is worth noting that the discovery years for our three unclassified SNe range from 1901--1959, when both spectroscopic and localization data were more uncertain than for modern SNe.

Our calculations thus show that when a SN appears coincident with an \ion{H}{2} region \textit{it almost always also appears coincident with CO~(2--1) emission}. Note that previous work on the CO-H$\alpha$ correlation has shown a significant de-correlation between these two tracers at high resolution \citep[e.g., see][]{Schruba2010,Kruijssen2019,Schinnerer2019,Pan2022,Leroy2023}. Those observations helped to establish the picture described above and suggest an important role for pre-SN feedback in clearing gas \citep[e.g.,][]{Chevance2020,Chevance2022}. The prevalence of CO detections coincident with SNe and \ion{H}{2} regions in our results may seem to contradict this result. However, we note a few points. First, $59\%$ of our SNe are \textit{not} detected coincident with \ion{H}{2} regions and there does not appear to be any statistically significant excess of SNe associated with CO outside \ion{H}{2} regions. So in good agreement with \citet{MaykerChen2023a}, many SNe do appear unassociated with recent star formation or molecular gas. Second, our control calculation suggests that to some degree the correlation that we do see simply reflects that we are capturing SNe in higher SFR parts of galaxies. Finally, we note that the resolution used for this calculation is still coarse. As we saw in \S \ref{sec:Zooms}, \ion{H}{2} regions are still small compared to the MUSE resolution, and the same is likely true of the molecular gas and ALMA. As a result, many of the joint \ion{H}{2} region-CO detections likely reflect complex regions with multi-generational star formation blended together by the $\sim 50{-}150$~pc resolution of the VLT and ALMA. Even with these caveats, the result here seems intriguing to us and worth investigation using higher resolution CO as well as H$\alpha$ data.

\subsection{H\alphaforsec\ and the Balmer Decrement}
\label{sec:BD}

\begin{figure}[ht!]
    \centering 
    \includegraphics[width=0.9\linewidth]{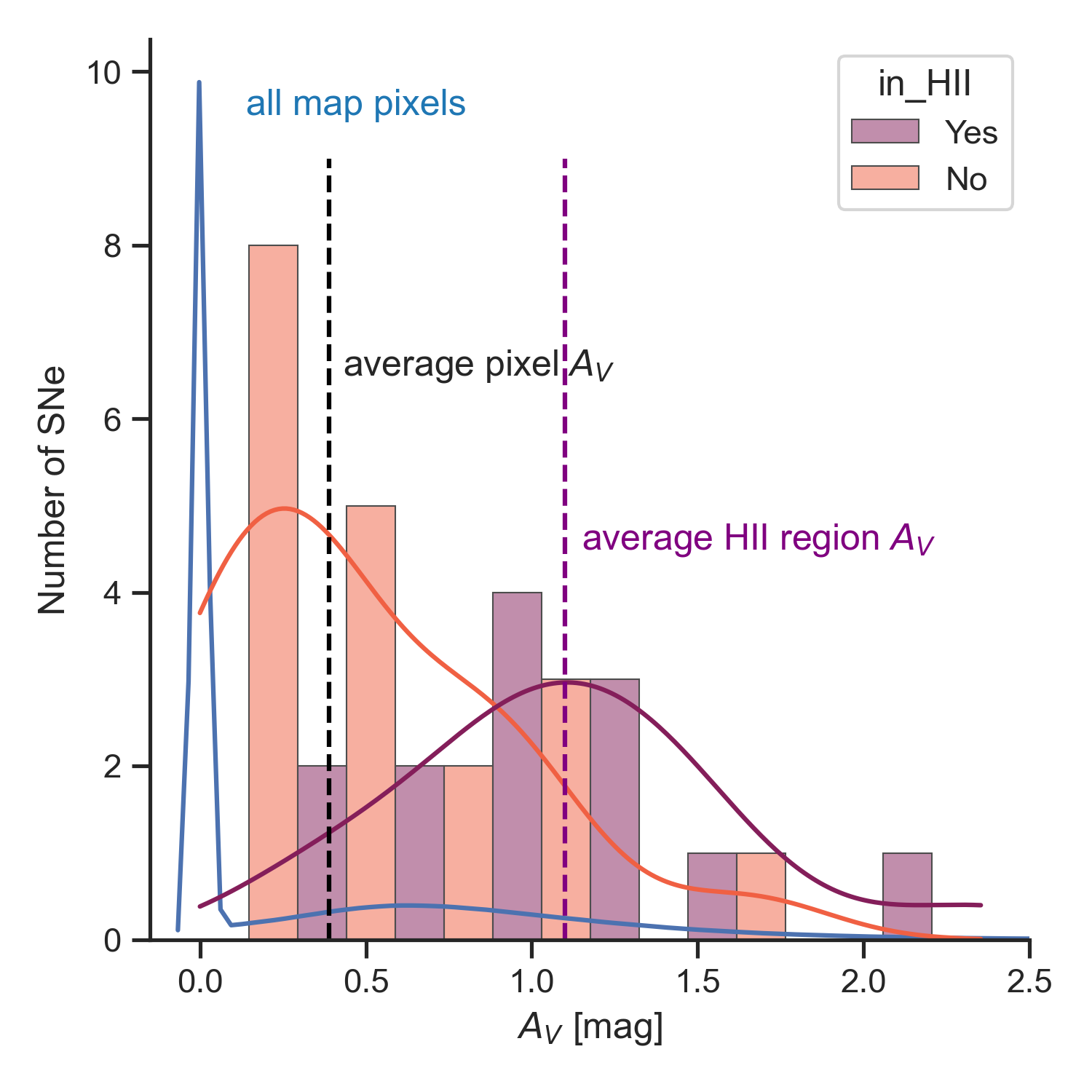}
\caption{$A_{V}$ values estimated by \citet{Belfiore2022,Belfiore2023} based on the Balmer decrement at each SN site. Lavender bars mark the distribution of $A_V$ for SNe that are coincident with \ion{H}{2} regions. Salmon bars show extinction for SNe not coincident with \ion{H}{2} region and are shifted to the right by one bar width. The KDE of the \ion{H}{2} and non-\ion{H}{2} populations are drawn in lavender and salmon lines respectively. The KDE for all pixels across all galaxies in our sample is shown by a blue line and the mean for all pixels and all \ion{H}{2} regions (with or without SNe) are shown by vertical dashed lines. The average extinction value for all pixels in all maps is $A_V \approx 0.4$~mag, while the average value for all pixels associated with \ion{H}{2} regions is $A_V \approx 1.1$~mag.}
\label{fig:ExtHist}
\end{figure}

The PHANGS-MUSE maps capture the extinction towards the ionized gas along each line of sight via the H$\beta$/H$\alpha$ ratio, the Balmer decrement. This allows us to assess the degree to which SNe preferentially occur in high extinction regions. This local extinction may impact the observability of SNe in surveys of more distant systems. Because gas and dust are mixed, the extinction also gives an alternative probe of the degree to which SNe explode near high column density gas. To test this, we measure extinction, expressed as $A_V$ values, based on the Balmer decrement along the line-of-sight towards each of our SNe. We adopt the calculations from \citet{Belfiore2023, Belfiore2022}, which adopts $R_{\rm V} = 3.1$ and an \citet{Odonnell1994} extinction curve. Note that this measurement captures the extinction towards the ionized gas along the line of sight (assuming a foreground screen geometry), which may or may not be identical to the extinction towards the SN itself.

In Figure~\ref{fig:ExtHist}, we plot the resulting histograms of extinction, expressed as $A_V$, towards SN sites, separating those coincident with an \ion{H}{2} region from those outside \ion{H}{2} regions. For comparison, we also plot the KDE of all pixels in the MUSE maps for our targets. The figure shows that our SNe occur in regions with extinction values ranging from $A_V = 0{-}2.35$~mag. The sites that are not coincident with \ion{H}{2} regions tend to have lower extinction, median $A_V=$ 0.33~mag with a $16{-}84\%$ range of 0--0.92~mag\footnote{In the \citet{Belfiore2022,Belfiore2023} maps, regions without detected H$\beta$ are set to $A_V = 0$~mag. The diffuse gas has lower overall intensity than \ion{H}{2} regions so some of these very low $A_V$ may be biased low due to the faintness of the lines.}, than the SN sites associated with \ion{H}{2} regions, which have median $A_V=$ 1.03~mag with a $16{-}84\%$ range of $0.75-1.47$~mag. These values resemble typical values found in the MUSE maps overall. In those maps a mean $A_V$ outside \ion{H}{2} regions is $A_V \approx 0.4$~mag and the mean $A_V$ associated with nebular regions is $A_V \approx 1.1$~mag. The lower extinction for non-\ion{H}{2} regions is well-known \citep[e.g., Eqn 9 of][]{Calzetti2001}.

These results indicate that SNe appear associated with typical extinctions, providing no strong evidence that there is a large population of deeply embedded SNe. We note that, because we study very nearby galaxies (compared to the distance out to which SNe tend to be detected in modern broader SN searches), it would be reasonable to expect even quite embedded SNe to be detected in these targets. Despite this, the highest extinction value associated with any SN in our sample is $A_V =$ 2.35~mag. 

The fact that we do not appear to find evidence for high extinction towards SNe in normal star-forming galaxies agrees with previous works that show the extinctions around SNe II to be generally small \citep[e.g.,][]{Pejcha2015}. We note that the evidence for significant populations of hidden or high-extinction SNe comes primarily from studying U/LIRG or starburst systems \citep[e.g.,][]{Fox2021}. In those systems, star-forming regions themselves are also found associated with high extinction and primarily visible in the radio or infrared \citep[e.g.,][]{Sanders1996}. Thus a more general phrasing of our results might be: our measurements support the idea that SNe occur at typical extinctions for their host galaxies, and are not particularly concentrated towards the highest or lowest $A_V$.

The modest extinctions that we measure towards SNe also support the idea of significant pre-SN clearing of material. The typical extinction of $A_V \sim 1$~mag that we find towards \ion{H}{2} regions with SNe corresponds to $\sim 20$~M$_\odot$~pc$^{-2}$ for a typical dust-to-gas ratio. For comparison, the surface densities associated with star formation tend to be $\sim 100$~M$_\odot$~pc$^{-2}$ \citep[e.g.,][]{Kennicutt2012}. Thus, also judging by extinction, most of the SNe in our sample appear to occur in regions where the progenitor has separated from its natal cloud. 

%%%%%%%%%%%%%%%%%%%%%%%%%%   SECTION 3.6  %%%%%%%%%%%%%%%%%%%%%%%%%%
\subsection{Additional Diagnostics}
\label{sec:Additional}

\begin{figure}
\centering
\includegraphics[width = 1.0\linewidth]{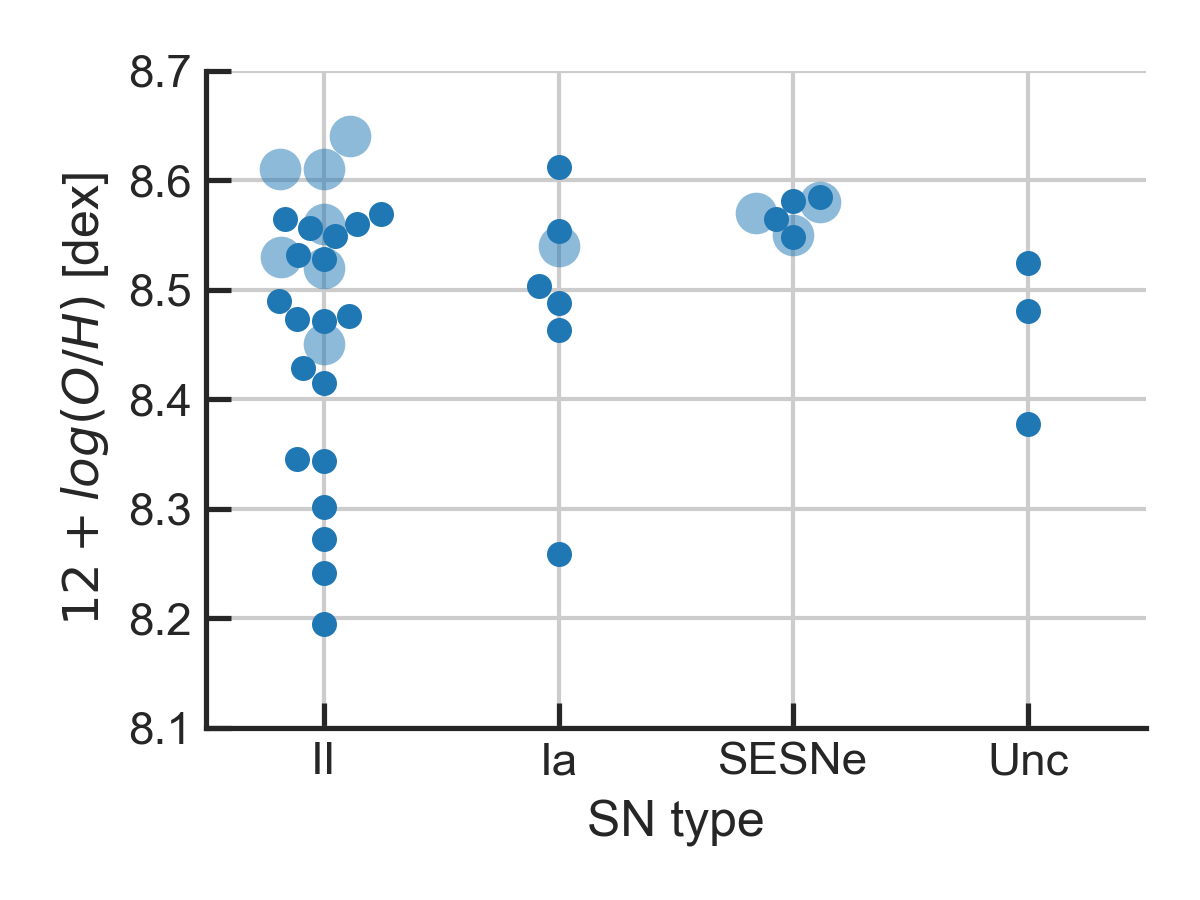}
\caption{Estimated metallicity values at each SN site based on \citet{Groves2023}. The small circles plot estimates of metallicity based on radial gradients. The large, transparent circles show estimates for the specific \ion{H}{2} region coincident with the SN. Both estimates use the Scal system. We separate the sample by SN type. \label{fig:met}}
\end{figure}

The \citet{Groves2023} catalog provides a treasure trove of information that can be used to further characterize the properties of ionized gas. We note several of these additional diagnostics for each of our SN sites in Table~\ref{tab:supernovaeTable}:

\begin{enumerate}

\item \textit{BPT classifications ([NII], [SII], [OI])}: \\The BPT emission line diagnostics \citep{Baldwin1981} are used to classify emission as coming from either star-forming regions, active galactic nuclei (AGN), low-ionization nuclear emission line regions (LINERs), or a composite of more than one \citep[for more information see \S 4.2 in][]{Groves2023}. We record the BPT emission line diagnostics for each of our SNe that occur along the line-of-sight to a nebular region. Of the 13 SNe associated with cataloged nebulae, all but two have all BPT diagnostics consistent with arising from star forming regions (i.e., reported as values of 0.0 in \citealt{Groves2023}). SN1997bs \& SN 2019ehk have two out of three emission line diagnostics consistent with star formation with the remaining line diagnostic indicative of a S/N ratio lower than 5 (non-informative, labeled as -1.0). This indicates that these are likely \ion{H}{2} regions but also might have an additional source of emission within the nebula.

\item \textit{Metallicity (12 + log (O/H))}: \\
In Figure~\ref{fig:met} we plot the estimated metallicity at each SN site. These estimates use the metallicity reported in \cite{Groves2023} based on ``Scal'' metallicity estimates \citep{Pilyugin:2016} for individual \ion{H}{2} regions. We also plot the specific metallicity values estimated at individual \ion{H}{2} regions for the subset of SNe within an \ion{H}{2} region.
% The metallicity dependence of SN progenitors of Ib/c etc. is a much less understood issue than the impact of their ages/masses (e.g. metallicity has a strong effect on the strengths of winds, which can affect Type I vs II statistics). 

\item \textit{Velocity dispersion measured from ionized gas ($\sigma$)}:\\
The velocity dispersion estimated from the ionized gas may give clue as to whether H$\alpha$ emission is coming from an \ion{H}{2} region or shocks, e.g., driven by winds or even the SN itself due to late-time CSM interactions.
We therefore also record the velocity dispersion ($\sigma$) of the H$\alpha$ emission at each SN site. None of our SN sample show signs of broadening with $\sigma \gg 100$~km~s$^{-1}$. 

\end{enumerate}

The nebulae coincident with our SNe appear consistent with normal \ion{H}{2} regions. The metallicities that we record are informational and potentially of use in broader future studies, such as determining the metallicity dependence on the rates of individual SN types \citep[e.g.,][]{Pessi2023}. Our own sample lacks a wide range in metallicity and the heterogeneous nature of our SN compilation makes it poorly suited to study SN rates.

%------------------- Discussion ----------------------%
\section{Discussion and Summary}
\label{sec:Discussion}

We have analyzed the coincidence of recent ($\lesssim 125$~yr old) supernovae (SNe) with H$\alpha$ emission in the PHANGS--MUSE survey \citep{Emsellem2022}. Within the $19$ PHANGS--MUSE targets, we identify $32$~SNe in $10$ galaxies that lie within the footprint of the MUSE observations and meet our criteria for being included within our sample.

\begin{enumerate}
\setcounter{enumi}{0}
\item We find that $41\%$ (13/32) of the SNe within the PHANGS--MUSE footprint occur coincident with one of the \ion{H}{2} regions identified by \citet[][]{Groves2023} (see \S \ref{sec:HIIregions} \& Table~\ref{tab:types}). The majority of these SNe (11/13) are also coincident with CO~(2--1) emission (\S \ref{sec:HACO}).
\end{enumerate}

We construct a series of models to test how much of this SNe-\ion{H}{2} region overlap may result from chance alignment (\S \ref{Analysis}).

\begin{enumerate}
\setcounter{enumi}{1}
\item Based on controls using the near-IR emission from the galaxy or considering random placement within a $500{-}1,000$~pc-sized region around the SNe, many of the SNe found coincident with \ion{H}{2} regions are likely to be associated by chance. Overall, by chance alone we would expect $\approx 1/3$ of SNe to appear associated with \ion{H}{2} regions for our targets and the MUSE resolution. This reflects that core collapse SNe tend to occur in regions of galaxies with a high density of star formation activity (\S \ref{sec:HIIregions}).

\item Examining the SN locations in new high resolution HST narrowband H$\alpha$ imaging (Chandar, Barnes et al. in prep.) confirms that SNe tend to explode \textit{near} but not directly \textit{in} \ion{H}{2} regions traced by H$\alpha$ (\S \ref{sec:Zooms}).

\item After accounting for this chance overlap, our best estimate is that $\approx 19.2 \pm 10.4\%$ of CCSNe occur within and are associated with an \ion{H}{2} region (\S \ref{sec:HIIregions}). Consistent with other recent work, this low percentage implies a large role for stellar winds, photoionized gas pressure, and radiation pressure (i.e., ``pre-SN'' feedback) in clearing gas away from young stellar populations. It also implies that SNe are more likely to explode into lower density regions and thus affect a larger physical area \citep[e.g.,][]{Chevance2020,MaykerChen2023a,Sarbadhicary2023}.
\end{enumerate}

We validate this result by analyzing intensity statistics and the distance from each SN to the nearest \ion{H}{2} region (\S \ref{sec:HIIregionsDist} and \ref{sec:HaEmission}). SNe Ia, which originate from older progenitors, show a closer association with the older stellar population compared to core collapse SNe. By contrast, stripped envelope SNe (SNe Ib/c), which are believed to originate from high mass stars, show the most direct association with H$\alpha$ emission and \ion{H}{2} regions out of our sample (\S \ref{sec:types}). These results by type are in good agreement with previous lower resolution work on larger samples \citep[including][]{James2006,Anderson2012,Crowther2013,Galbany2014}.

In addition to H$\alpha$ intensity, we examine extinction and other properties of the nebular regions associated with SNe.

\begin{enumerate}
\setcounter{enumi}{4}
\item The Balmer decrement implies extinctions towards the ionized gas near SNe of $A_V \approx 0{-}2.35$~mag. We find no strong differences between the overall $A_V$ values in the PHANGS--MUSE maps and those at the sites of SNe. This appears consistent with modest measured $A_V$ towards core collapse SNe in normal star-forming galaxies (\S \ref{sec:BD}).
\end{enumerate}

Thanks to ASAS-SN, ATLAS, ZTF, and the upcoming \textit{Rubin} LSST, we are in the era of high completeness SNe discovery towards nearby galaxies. Therefore we expect the kind of detailed, high physical resolution studies presented here and in \citet{MaykerChen2023a} to become increasingly possible and informative. We note two conclusions relevant to such next steps:

\begin{enumerate}
\setcounter{enumi}{5}
\item High physical resolution, of order $10$~pc, is needed to isolate SNe within individual clouds or \ion{H}{2} regions. Given that core collapse SNe often occur in complex regions of galaxies with active star formation, coarser resolution risks significant chance accidental alignment. This is also of order the resolution needed to resolve the likely cooling radius (and so the zone of influence) of SN explosions \citep[e.g.,][]{Kim2015,Martizzi2015}.
\item To achieve such high resolution, correspondingly high accuracy in the localization of SNe is required. Practically, to make best use of space telescopes or ALMA for such studies, SNe must be localized to better than $\pm$0\farcs1.
\end{enumerate}

With these caveats in mind, the future in this area looks bright. JWST, HST, ALMA, and soon \textit{Euclid} and \textit{Roman} offer amazing prospects to localize SNe in the nearest galaxies and better understand the impact and origin of these explosions.

%-------------------------ACKNOWLEDGEMENTS-----------------------------%
\section*{Acknowledgments}

NMC thanks the Ohio State University's Galaxy and Supernova groups, including Christopher Kochanek, for useful discussions at several stages of the project. This work was carried out as part of the PHANGS collaboration.

Support for this work was provided by the NSF through award SOSP SOSPADA-010 from the NRAO, which supported the work of NMC. The work of NMC and AKL was partially supported by the National Science Foundation (NSF) under Grants No.~1615105, 1615109, and 1653300.

The work of AKL is partially supported by the National Aeronautics and Space Administration (NASA) under ADAP grants NNX16AF48G and NNX17AF39G.
KG is supported by the Australian Research Council through the Discovery Early Career Researcher Award (DECRA) Fellowship (project number DE220100766) funded by the Australian Government. 

KK, JEMD, and OVE gratefully acknowledge funding from the Deutsche Forschungsgemeinschaft (DFG, German Research Foundation) in the form of an Emmy Noether Research Group (grant number KR4598/2-1, PI Kreckel) and the European Research Council’s starting grant ERC StG-101077573 (“ISM-METALS"). 

M.C. gratefully acknowledges funding from the Deutsche Forschungsgemeinschaft (DFG, German Research Foundation) through an Emmy Noether Research Group (grant number CH2137/1-1). COOL Research DAO is a Decentralised Autonomous Organisation supporting research in astrophysics aimed at uncovering our cosmic origins.

This paper makes use of the PHANGS--MUSE data, based on observations collected at the European Southern Observatory under ESO programmes 094.C-0623 (PI: Kreckel), 095.C-0473,  098.C-0484 (PI: Blanc), 1100.B-0651 (PHANGS-MUSE; PI: Schinnerer), as well as 094.B-0321 (MAGNUM; PI: Marconi), 099.B-0242, 0100.B-0116, 098.B-0551 (MAD; PI: Carollo) and 097.B-0640 (TIMER; PI: Gadotti). 

This paper makes use of the following ALMA data, which have been processed as part of the PHANGS--ALMA CO~(2--1) survey: \\
\noindent 
ADS/JAO.ALMA\#2012.1.00650.S\\ 
ADS/JAO.ALMA\#2013.1.00803.S\\
ADS/JAO.ALMA\#2013.1.01161.S\\
ADS/JAO.ALMA\#2015.1.00121.S\\
ADS/JAO.ALMA\#2015.1.00782.S\\
ADS/JAO.ALMA\#2015.1.00925.S\\
ADS/JAO.ALMA\#2015.1.00956.S\\
ADS/JAO.ALMA\#2016.1.00386.S\\
ADS/JAO.ALMA\#2017.1.00392.S\\
ADS/JAO.ALMA\#2017.1.00766.S\\
ADS/JAO.ALMA\#2017.1.00886.L\\
ADS/JAO.ALMA\#2018.1.00484.S\\
ADS/JAO.ALMA\#2018.1.01321.S\\
ADS/JAO.ALMA\#2018.1.01651.S\\
ADS/JAO.ALMA\#2018.A.00062.S\\
ADS/JAO.ALMA\#2019.1.01235.S\\
ADS/JAO.ALMA\#2019.2.00129.S\\
ALMA is a partnership of ESO (representing its member states), NSF (USA), and NINS (Japan), together with NRC (Canada), NSC and ASIAA (Taiwan), and KASI (Republic of Korea), in cooperation with the Republic of Chile. The Joint ALMA Observatory is operated by ESO, AUI/NRAO, and NAOJ. The National Radio Astronomy Observatory (NRAO) is a facility of NSF operated under cooperative agreement by Associated Universities, Inc (AUI).

This research is based on observations made with the NASA/ESA Hubble Space Telescope obtained from the Space Telescope Science Institute, which is operated by the Association of Universities for Research in Astronomy, Inc., under NASA contract  NAS 5–26555. These observations are associated with program 15654.

This work has made use of the Transient Name Server, provided by the IAU supernova working group, last accessed on 2023 December 30, which included observations and metadata for over 140,000 astronomical transients.

This work acknowledges the Open Supernova Catalog (OSC; \citealt{Guillochon2017}), last accessed on 2022, January 26th, which included observations and metadata for $\sim$80,000 SNe \& SNRs.

This work has made use of SAO/NASA Astrophysics Data System
%\footnote{\url{http://www.adsabs.harvard.edu}}
, the NASA/IPAC Extragalactic Database (NED), which is operated by the Jet Propulsion Laboratory, California Institute of Technology, under contract with the National Aeronautics and Space Administration, the SIMBAD database, operated at CDS, Strasbourg, France, and the Central Bureau for Astronomical Telegrams (CBAT) and the International Astronomical Union Circulars (IAUC) catalog %\footnote{\url{http://www.cbat.eps.harvard.edu/services/IAUC.html}}
, operated at Harvard College Observatory, Cambridge, Massachusetts. 

This work has utilized the following software: Jupyter \citep{Kluyver2016}, Astropy \citep{Astropy2013, Price-Whelan2018}, Pandas \citep{McKinney2010}, NumPy \citep{VanDerWalt2011, Harris2020}, SciPy \citep{Virtanen2020}, Seaborn \citep{Waskom2017}, Matplotlib \citep{Hunter2007}, \& GitHub \citep{github}.

%%%%%%%%%%%%%%%%%%%%%%   BIBLIOGRAPHY   %%%%%%%%%%%%%%%%

\bibliography{mybib.bib}

%\clearpage{}

\appendix
\label{appendix}

% reset counters
\renewcommand{\thefigure}{A\arabic{figure}}
\setcounter{figure}{0}
\renewcommand{\thetable}{A\arabic{table}}
\setcounter{table}{0}
\renewcommand{\theequation}{A\arabic{equation}}
\setcounter{equation}{0}

%------------------ Galaxies Table --------------------%

\begin{deluxetable*}{lcccccc}[]
\tabletypesize{\scriptsize}
\tablecaption{Supernovae in MUSE Galaxies}
\label{tab:galaxyTable}
\tablehead{
\colhead{Galaxy} &
\colhead{Supernova} & 
\colhead{Type} & 
\colhead{R.A.} &
\colhead{Dec} &
\colhead{In Sample} &
\colhead{Reference}}
\startdata
NGC0628 & SN2013ej & II & 24.2007 & 15.7586 & - & \citet{Leonard2013} \\
NGC0628 & SN2019krl & IIn/LBV & 24.2068 & 15.7795 & - & \citet{Andrews2021} \\
NGC1087 & SN1995V & II & 41.6115 & - 0.4988 & \checkmark & \citet{Evans1995} \\
NGC1300 & SN2022acko & II & 49.9125 & -19.3952 & \checkmark & \citet{Li2022} \\
NGC1365 & SN1957C & Unclassified & 53.3835 & -36.1177 & \checkmark & HAC 1383\\
NGC1365 & SN1983V & Ib & 53.3819 & -36.1486 & \checkmark & \citet{Wheeler1987} \\
NGC1365 & SN2001du & II & 53.3713 & -36.1421 & \checkmark & \citet{Smartt2001} \\
NGC1365 & SN2012fr & Ia & 53.4006 & -36.1268 & \checkmark & \citet{Klotz2012} \\
NGC1433 & SN1985P & II & 55.5264 & -47.21 & \checkmark & \citet{Kirshner1985} \\
NGC1566 & ASASSN-14ha & II & 65.0059 & -54.9381 & \checkmark & \citet{Arcavi2014} \\
NGC1566 & SN2010el & Ia-02cx & 64.9951 & -54.944 & \checkmark & \citet{Bessell2010} \\
NGC1566 & SN2021aefx & Ia & 64.9725 & -54.9481 & \checkmark & \citet{Valenti2021} \\
NGC1672 & SN2017gax & Ib/c & 71.4561 & -59.2451 & - & \citet{Jha2017} \\
NGC1672 & SN2022aau & II & 71.424 & -59.2454 & \checkmark & \citet{Siebert2022} \\
NGC3627 & SN1973R & II & 170.0481 & 12.9977 & \checkmark & \citet{Ciatti1977} \\
NGC3627 & SN1989B & Ia & 170.058 & 13.0053 & \checkmark & \citet{Marvin1989} \\
NGC3627 & SN1997bs & IIn & 170.0593 & 12.9721 & \checkmark & \citet{Adams2015} \\
NGC3627 & SN2009hd & II & 170.0707 & 12.9796 & \checkmark & \citet{Kasliwal2009} \\
NGC3627 & SN2016cok & II P & 170.0796 & 12.9824 & \checkmark & \citet{Zhang2016} \\
NGC4254 & SN1967H & II & 184.7184 & 14.414 & \checkmark & \citet{Fairall1967} \\
NGC4254 & SN1972Q & II & 184.7107 & 14.4443 & \checkmark & \citet{Rosino1972} \\
NGC4254 & SN1986I & II & 184.7169 & 14.4123 & \checkmark & \citet{Pennypacker1986} \\
NGC4254 & SN2014L & Ic & 184.7029 & 14.4121 & \checkmark & \citet{Yamaoka2014} \\
NGC4303 & SN1926A & II & 185.4754 & 4.4934 & \checkmark & IAUC 111 \\
NGC4303 & SN1961I & II & 185.5018 & 4.4704 & \checkmark & \citet{Humason1962} \\
NGC4303 & SN1964F & II & 185.4698 & 4.4738 & \checkmark & IAUC 1868  \\
NGC4303 & SN1999gn & II P & 185.4876 & 4.4627 & \checkmark & \citet{Ayani1999} \\
NGC4303 & SN2006ov & II & 185.4804 & 4.488 & \checkmark & \citet{Blondin2006} \\
NGC4303 & SN2014dt & Ia Pec & 185.4899 & 4.4718 & \checkmark & \citet{Ochner2014} \\
NGC4303 & SN2020jfo & II P & 185.4602 & 4.4817 & \checkmark & \citet{Perley2020} \\
NGC4321 & SN1901B & I & 185.6971 & 15.8238 & \checkmark & \citet{Tsvetkov1993} \\
NGC4321 & SN1959E & I & 185.7454 & 15.817 & \checkmark & \citet{Porter1993} \\
NGC4321 & SN1979C & II & 185.7442 & 15.7978 & - & \cite{Carney1980} \\
NGC4321 & SN2006X & Ia & 185.7249 & 15.809 & \checkmark & \citet{Quimby2006} \\
NGC4321 & SN2019ehk & Ib & 185.7339 & 15.8261 & \checkmark & \citet{De2021} \\
NGC4321 & SN2020oi & Ic & 185.7289 & 15.8236 & \checkmark & \citet{Siebert2020} \\
\enddata
\end{deluxetable*}

%------------------ Supernovae Table --------------------%

\begin{deluxetable*}{lccccccccccclcc}[]
\tabletypesize{\scriptsize}
\tablecaption{Measurements from the Supernovae Sample}
\label{tab:supernovaeTable}
\tablehead{\colhead{Supernova} &
\colhead{Type} &
\colhead{Resolution} &
\colhead{Distance} &
\colhead{H$_{\alpha}$} &
\colhead{$\sigma$} &
\colhead{$A_V$} &
\colhead{\ion{H}{2}} &
\colhead{NII} &
\colhead{SII} &
\colhead{OI} &
\colhead{CO} &
\colhead{Metallicity} &
\colhead{$R_{\rm gal}$} &
\colhead{$R_{\rm eff}$}}
\startdata
ASASSN-14ha & II & 0.8 (80) & 17.69 & 621 & 57 & 0.86 & Yes (877) & 0.0 & 0.0 & 0.0 & Yes & 8.56 (8.61) & 0.05 & 1.31 \\
SN1901B & I & 1.16 (43) & 15.21 & 211 & 53 & 0.92 & No (-1) & - & - & - & Yes & 8.38 & 0.57 & 7.67 \\
SN1926A & II & 0.78 (73) & 16.99 & 82 & 67 & 0.0 & No (-1) & - & - & - & No & 8.34 & 0.35 & 8.42 \\
SN1957C & Unclassified & 1.15 (76) & 19.57 & 294 & 63 & 0.38 & No (-1) & - & - & - & No & 8.52 & 0.81 & 43.17 \\
SN1959E & I & 1.16 (43) & 15.21 & 361 & 55 & 1.07 & No (-1) & - & - & - & Yes & 8.48 & 0.3 & 3.98 \\
SN1961I & II & 0.78 (73) & 16.99 & 415 & 58 & 0.35 & Yes (373) & 0.0 & 0.0 & 0.0 & No & 8.3 (8.56) & 0.4 & 9.73 \\
SN1964F & II & 0.78 (73) & 16.99 & 29 & 59 & 0.0 & No (-1) & - & - & - & No & 8.49 & 0.16 & 3.85 \\
SN1967H & II & 0.89 (96) & 13.1 & 1435 & 63 & 1.65 & No (-1) & - & - & - & Yes & 8.47 & 0.16 & 4.24 \\
SN1972Q & II & 0.89 (96) & 13.1 & 362 & 56 & 0.78 & No (-1) & - & - & - & Yes & 8.27 & 0.43 & 11.34 \\
SN1973R & II & 1.05 (96) & 11.32 & 1118 & 59 & 1.17 & Yes (433) & 0.0 & 0.0 & 0.0 & Yes & 8.56 (8.52) & 0.29 & 4.39 \\
SN1983V & Ic & 1.15 (76) & 19.57 & 1485 & 66 & 0.95 & Yes (385) & 0.0 & 0.0 & 0.0 & Yes & 8.58 (8.57) & 0.48 & 25.77 \\
SN1985P & II & 0.91 (69) & 18.63 & 25 & 67 & 0.24 & No (-1) & - & - & - & No & 8.47 & 0.35 & 7.37 \\
SN1986I & II & 0.89 (96) & 13.1 & 515 & 62 & 0.52 & No (-1) & - & - & - & Yes & 8.48 & 0.15 & 4.08 \\
SN1989B & Ia & 1.05 (96) & 11.32 & 1257 & 66 & 0.81 & Yes (390) & 0.0 & 0.0 & 0.0 & Yes & 8.55 (8.54) & 0.17 & 2.58 \\
SN1995V & II & 0.92 (84) & 15.85 & 884 & 52 & 1.28 & Yes (731) & 0.0 & 0.0 & 0.0 & Yes & 8.24 (8.45) & 0.14 & 3.4 \\
SN1997bs & IIn & 1.05 (96) & 11.32 & 99 & 55 & 0.78 & Yes (961) & 1.0 & 0.0 & 0.0 & Yes & 8.56 (nan) & 0.24 & 3.66 \\
SN1999gn & II P & 0.78 (73) & 16.99 & 1285 & 61 & 0.41 & Yes (58) & 0.0 & 0.0 & 0.0 & Yes & 8.43 (8.64) & 0.24 & 5.76 \\
SN2001du & II & 1.15 (76) & 19.57 & 135 & 68 & 0.18 & No (-1) & - & - & - & No & 8.53 & 0.79 & 41.99 \\
SN2006X & Ia & 1.16 (43) & 15.21 & 37 & 60 & 0.54 & No (-1) & - & - & - & No & 8.5 & 0.24 & 3.17 \\
SN2006ov & II & 0.78 (73) & 16.99 & 331 & 68 & 0.33 & No (-1) & - & - & - & No & 8.41 & 0.26 & 6.19 \\
SN2009hd & II & 1.05 (96) & 11.32 & 3081 & 68 & 1.03 & Yes (396) & 0.0 & 0.0 & 0.0 & Yes & 8.56 (8.53) & 0.2 & 3.0 \\
SN2010el & Ia & 0.8 (80) & 17.69 & 38 & 68 & 0.26 & No (-1) & - & - & - & Yes & 8.49 & 0.13 & 3.39 \\
SN2012fr & Ia & 1.15 (76) & 19.57 & 9 & 68 & 0.0 & No (-1) & - & - & - & No & 8.61 & 0.31 & 16.54 \\
SN2014L & Ic & 0.89 (96) & 13.1 & 3366 & 63 & 1.47 & Yes (1309) & 0.0 & 0.0 & 0.0 & Yes & 8.53 (8.55) & 0.08 & 2.08 \\
SN2014dt & Ia Pec & 0.78 (73) & 16.99 & 90 & 71 & 0.28 & No (-1) & - & - & - & No & 8.46 & 0.19 & 4.69 \\
SN2016cok & II P & 1.05 (96) & 11.32 & 518 & 62 & 1.36 & No (-1) & - & - & - & Yes & 8.57 & 0.35 & 5.22 \\
SN2019ehk & Ib & 1.16 (43) & 15.21 & 104 & 76 & 1.47 & Yes (1722) & 1.0 & 0.0 & 0.0 & No & 8.55 (nan) & 0.12 & 1.59 \\
SN2020jfo & II P & 0.78 (73) & 16.99 & 220 & 57 & 0.92 & No (-1) & - & - & - & No & 8.35 & 0.35 & 8.37 \\
SN2020oi & Ic & 1.16 (43) & 15.21 & 5473 & 66 & 1.01 & Yes (1693) & 0.0 & 0.0 & 0.0 & Yes & 8.58 (8.58) & 0.02 & 0.28 \\
SN2021aefx & Ia & 0.8 (80) & 17.69 & 13 & 73 & 0.0 & No (-1) & - & - & - & No & 8.26 & 0.35 & 9.59 \\
SN2022aau & II & 0.96 (104) & 19.4 & 9695 & 76 & 2.35 & Yes (803) & 0.0 & 0.0 & 0.0 & Yes & 8.55 (8.61) & 0.05 & 1.32 \\
SN2022acko & II & 0.89 (83) & 18.99 & 34 & 66 & 0.3 & No (-1) & - & - & - & No & 8.19 & 0.38 & 5.35 \\
\enddata
\tablecomments{Resolution in arcseconds (pc), distance in Mpc, extinction-corrected H$_{\alpha}$ intensity in flux units of $10^{37} {\rm erg/s/kpc}^{2}$, $\sigma$ ($H_\alpha$ velocity dispersion, corrected for instrumental broadening) in units of km/s, $A_V$ derived using the Balmer decrement \citep{Groves2023}, adopting $R_{\rm V} = 3.1$ and an \citet{Odonnell1994} extinction curve, \ion{H}{2} regions identified in \citet{Groves2023}, \ion{N}{2}, \ion{S}{2}, \ion{O}{1} BPT emission line diagnostics taken from \cite{Groves2023}, CO~(2-1) detection assigned when signal-to-noise $\geq$ 3 from PHANGS--ALMA broad moment-0 maps \citep{Leroy2021}, metallicity in [12+log(O/H) (dex)] calculated using the  galactic metallicity gradient, determined using Scal metallicities, and the galactocentric radius of each SN (empirical values from \cite{Groves2023} when available), $R_{\rm gal}$ in kpc.}
\end{deluxetable*}

%------------------Zooms Figure -----------%

\begin{figure*}[]
\centering 
%\subfloat[]{%
\includegraphics[width=0.8\linewidth]{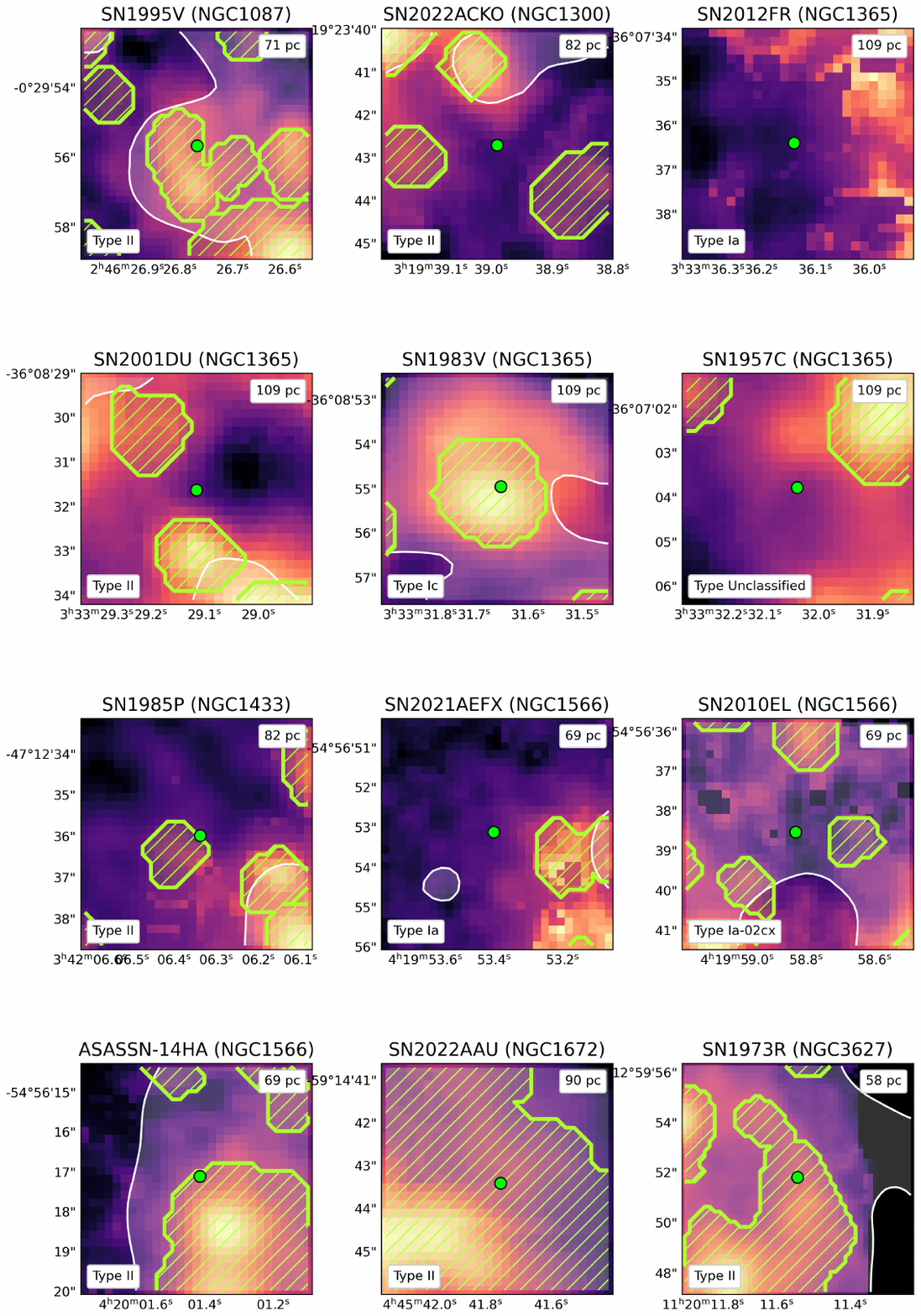}
%}\textit
\caption{\label{fig:1.a}%
\textit{PHANGS--MUSE H$\alpha$ emission cut-outs ($500 \times 500~pc$) centered on the locations of the SNe in our sample.} Each galaxy is plotted at its native resolution, listed in the top right corner. Lime contours enclose \ion{H}{2} regions identified by \citet{Groves2023}. SNe are marked with white circles. In the legend, SNe are labeled with their type classification. Each panel is oriented with the top of the figure as North, East is left}.
\end{figure*}

\begin{figure*}[]
\centering 
\includegraphics[width=0.8\linewidth]{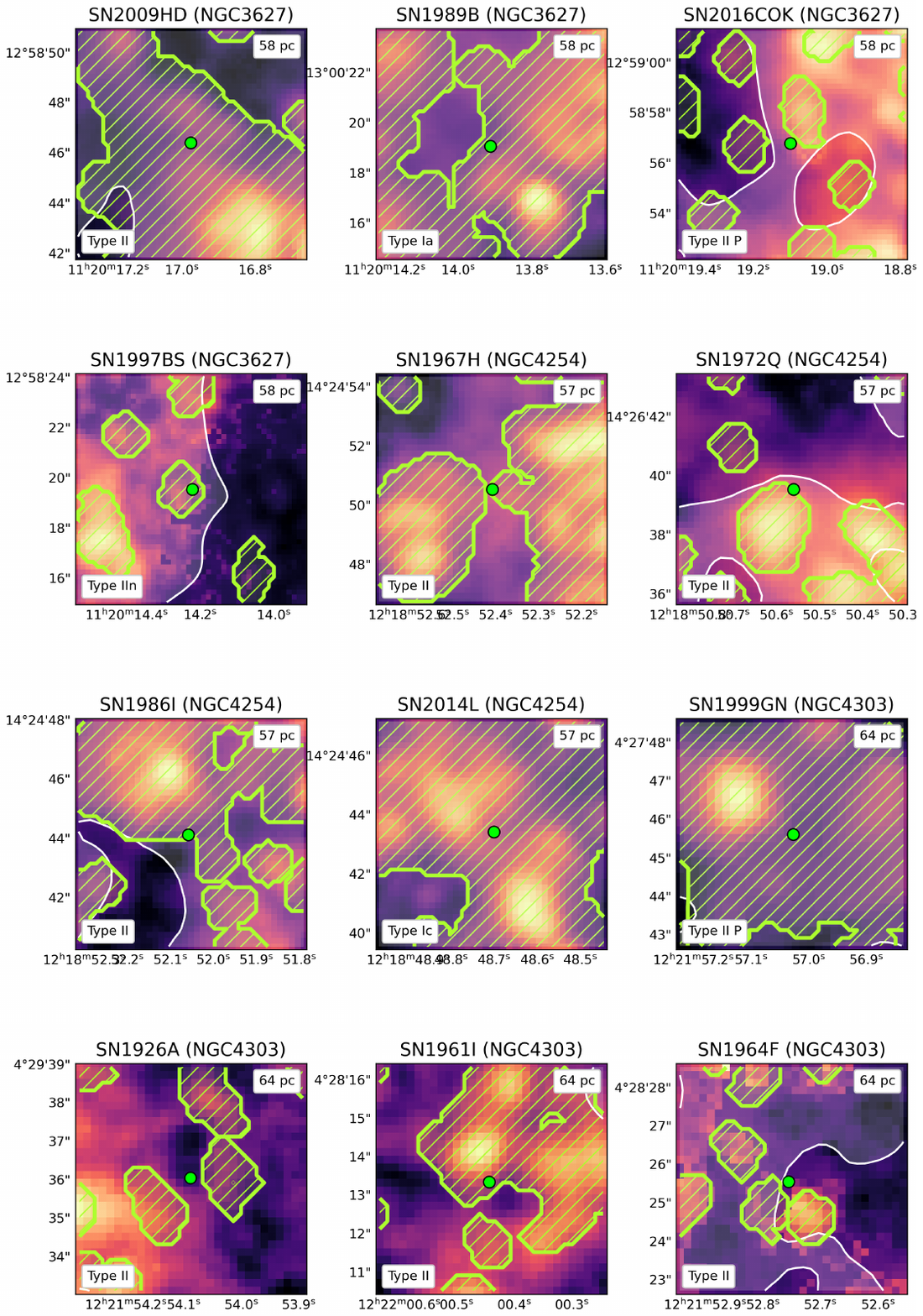}
\caption{\label{fig:1.b} Figure \ref{fig:1.a} continued.}
\end{figure*}

 \begin{figure*}[]
\centering 
\includegraphics[width=0.8\linewidth]{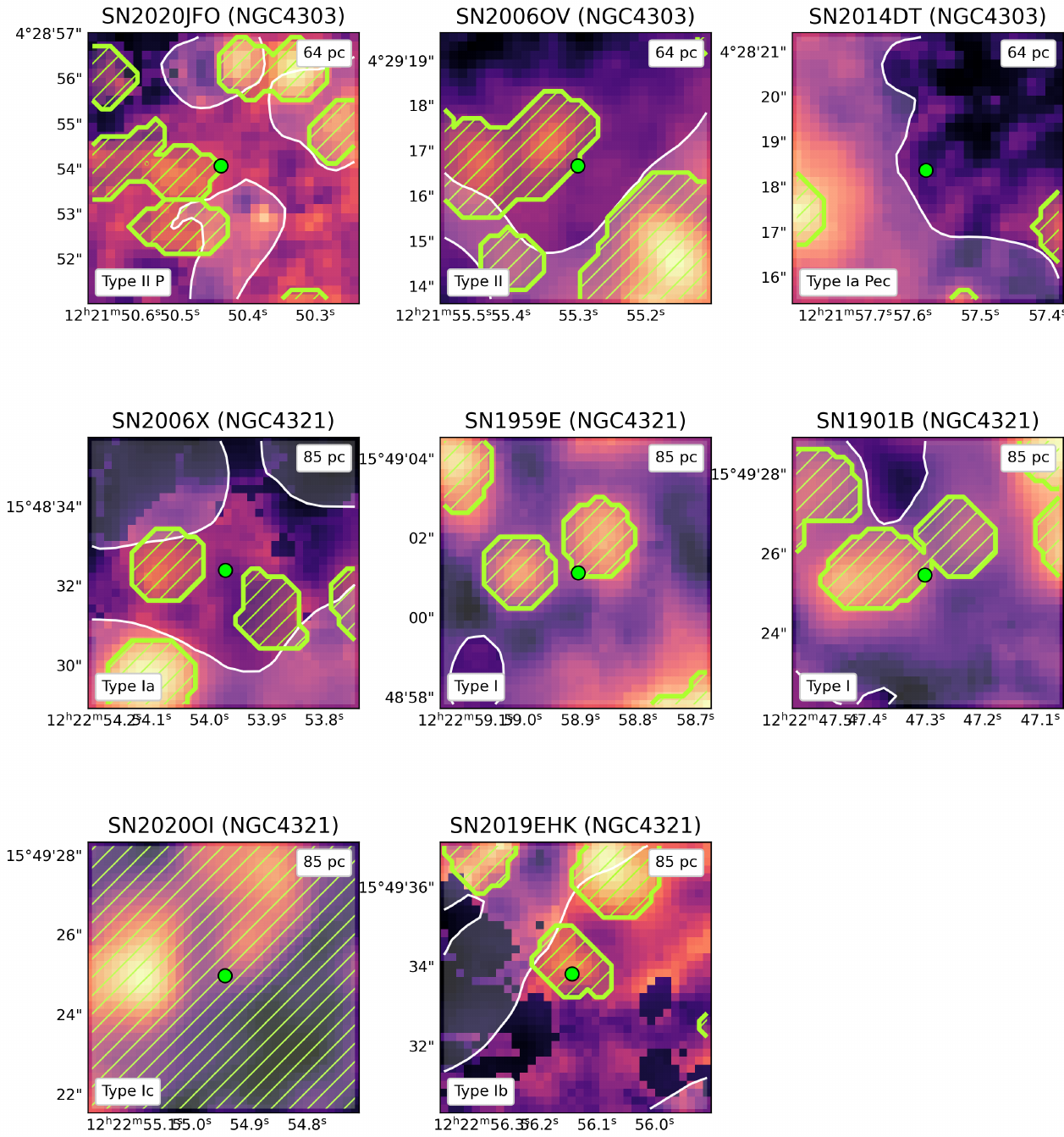}
%}
\caption{Figure \ref{fig:1.a} continued.}
\end{figure*}

\begin{figure*}
\centering
\includegraphics[width=0.8\textwidth]{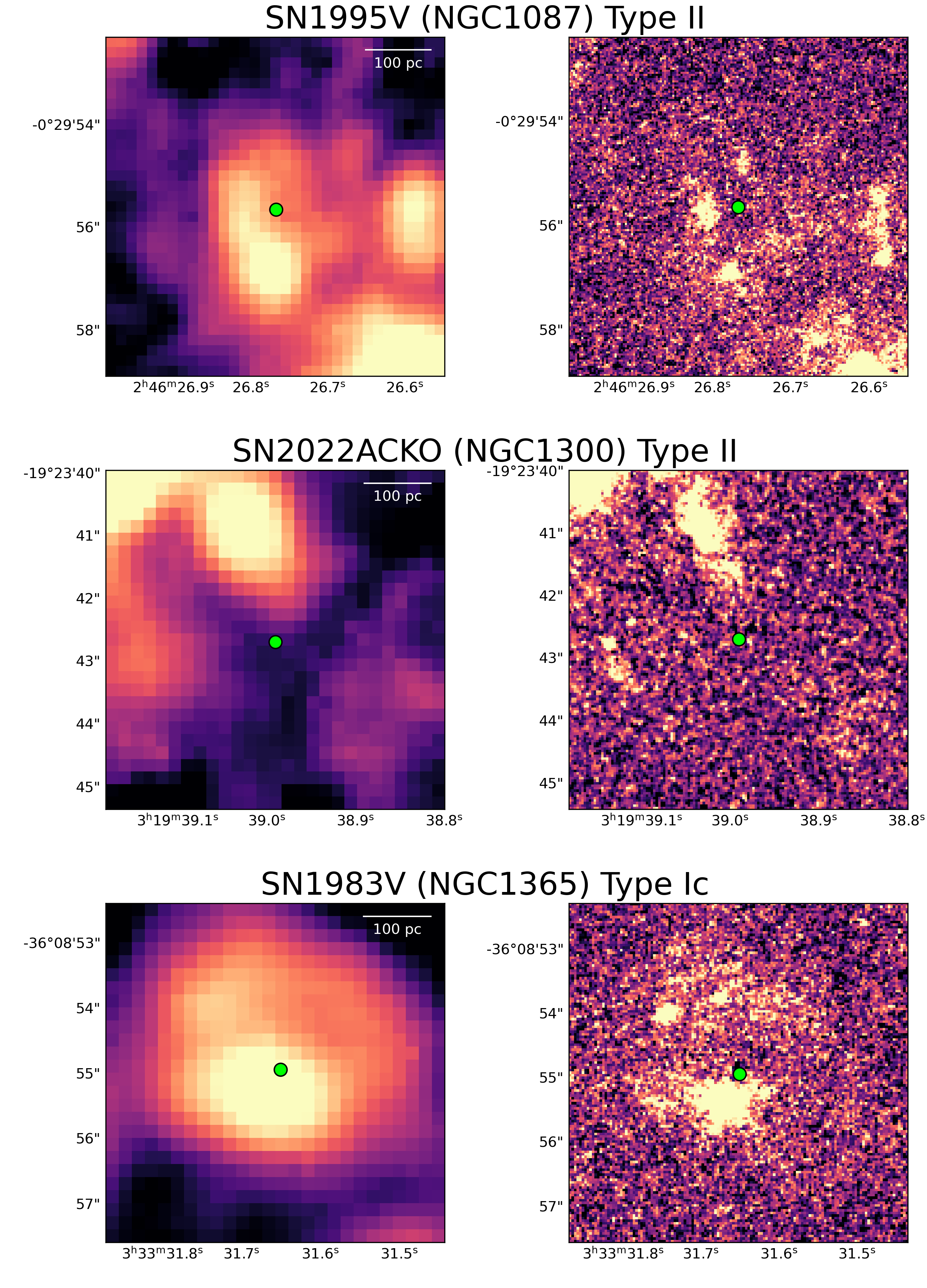} \caption{\textit{Comparison of PHANGS--MUSE (left panels) and HST (right panels) H$\alpha$ emission cut-outs ($500 \times 500~pc$) centered on the locations of the SNe in our sample.} SNe are marked with green circles. SNe are labeled with their host galaxy and type classification in the title of each subplot row. Each panel is oriented with the top of the figure as North, East is left \label{fig:HST}}
\end{figure*}
\begin{figure*}
\includegraphics[width=0.9\textwidth]{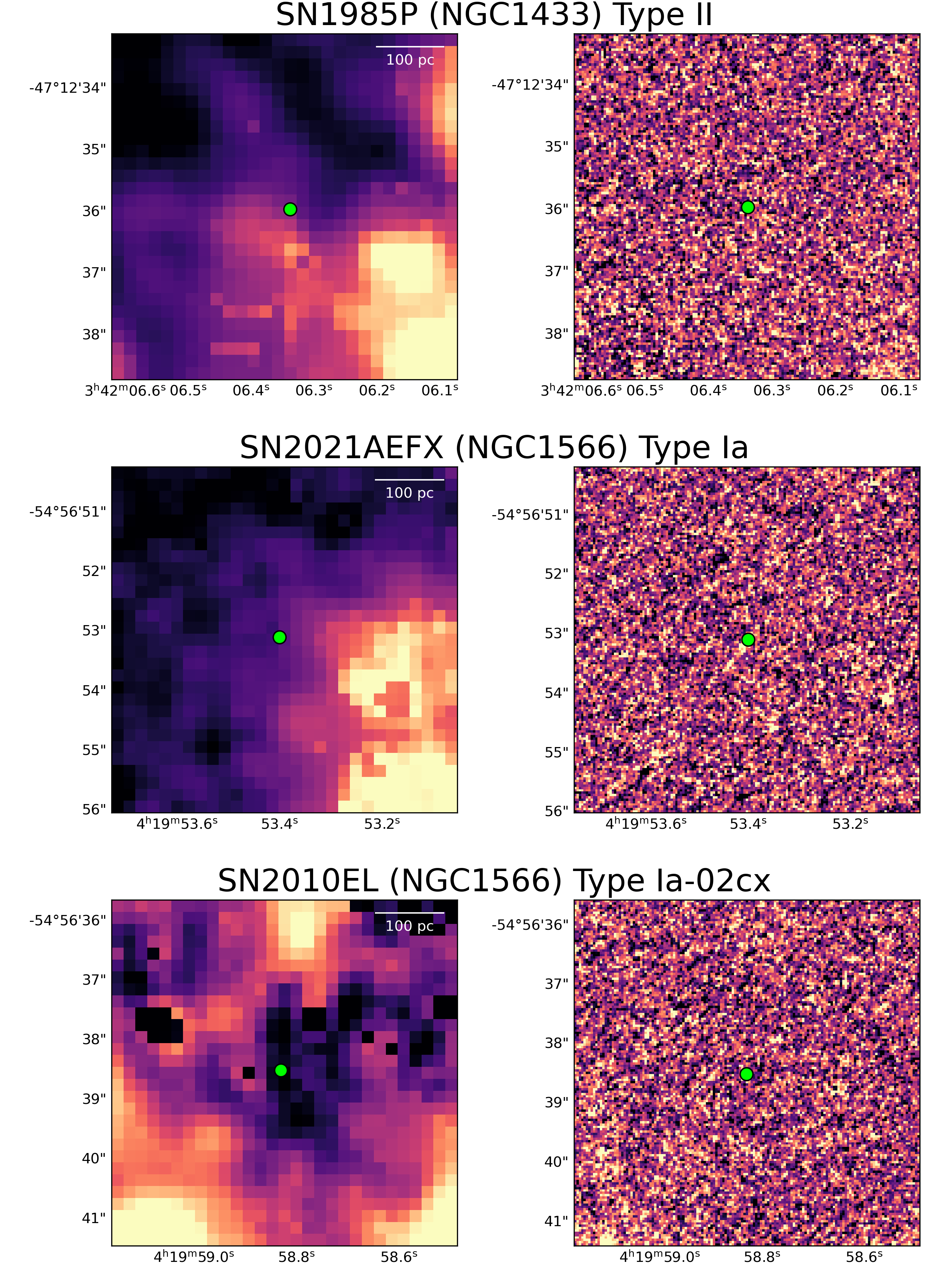}
\caption{Figure \ref{fig:HST} continued}
\end{figure*}
\begin{figure*}
\includegraphics[width=0.9\textwidth]{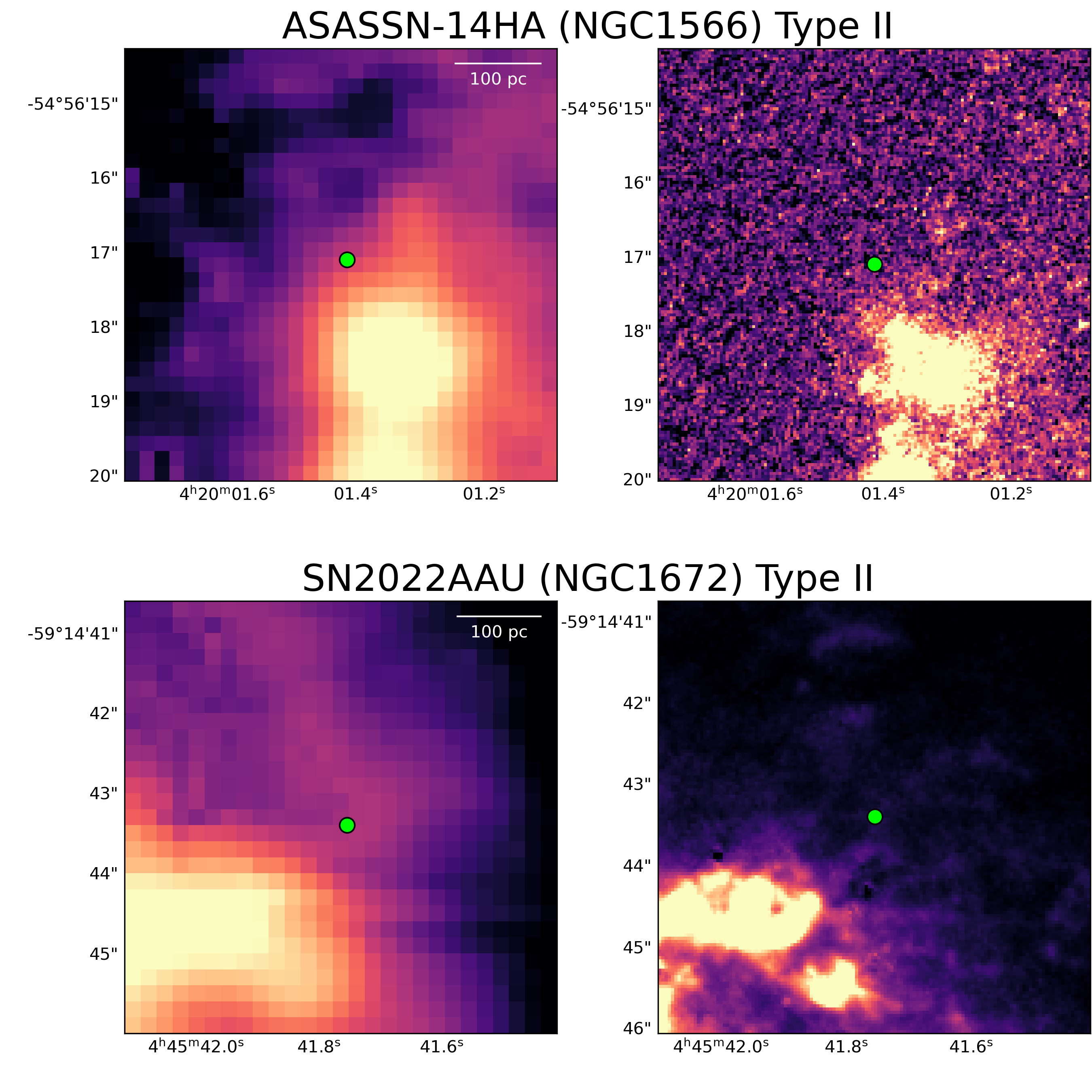}
\caption{Figure \ref{fig:HST} continued}
\end{figure*}
\begin{figure*}
\includegraphics[width=0.9\textwidth]{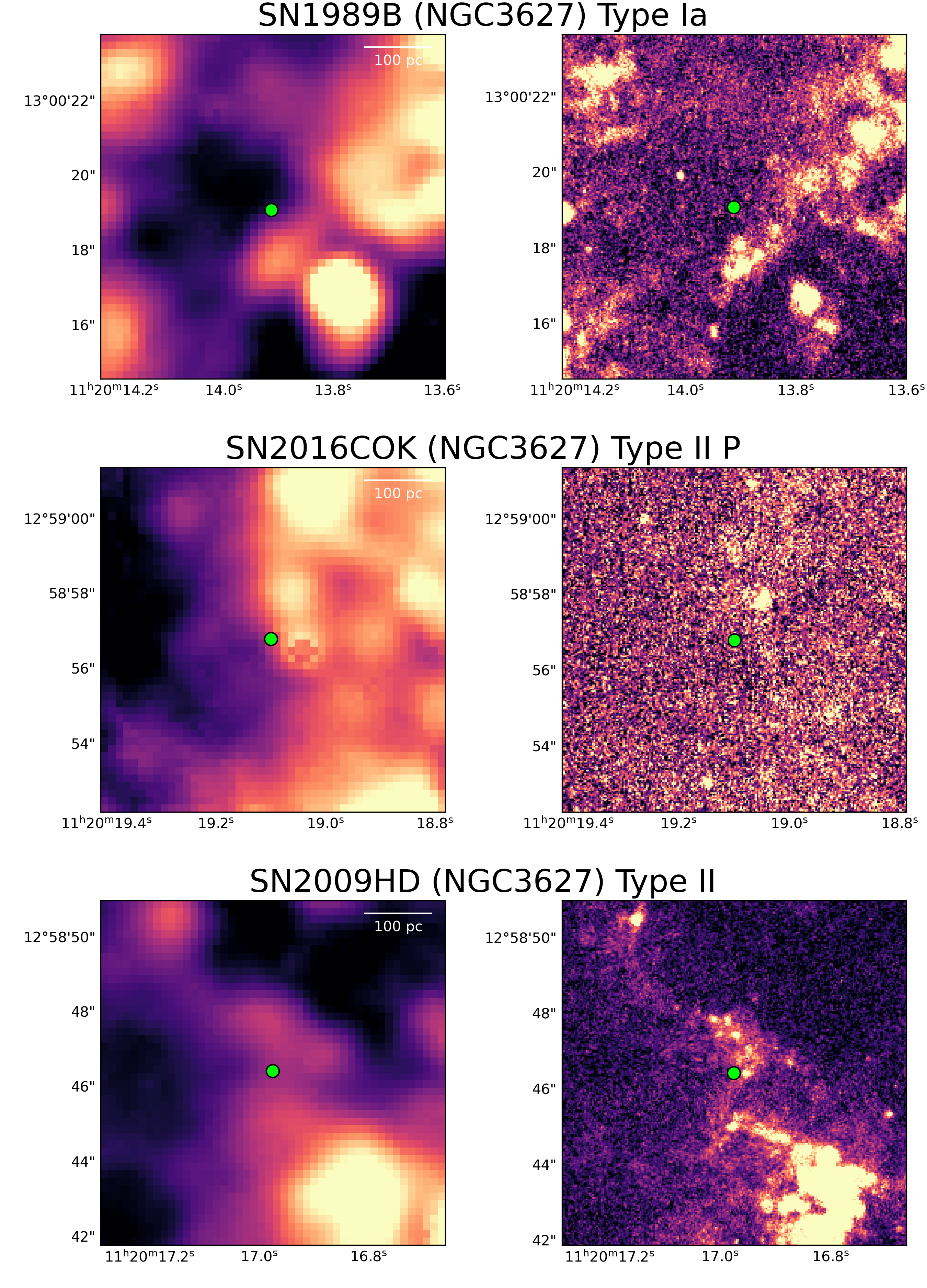}
\caption{Figure \ref{fig:HST} continued}
\end{figure*}
\begin{figure*}
\includegraphics[width=0.9\textwidth]{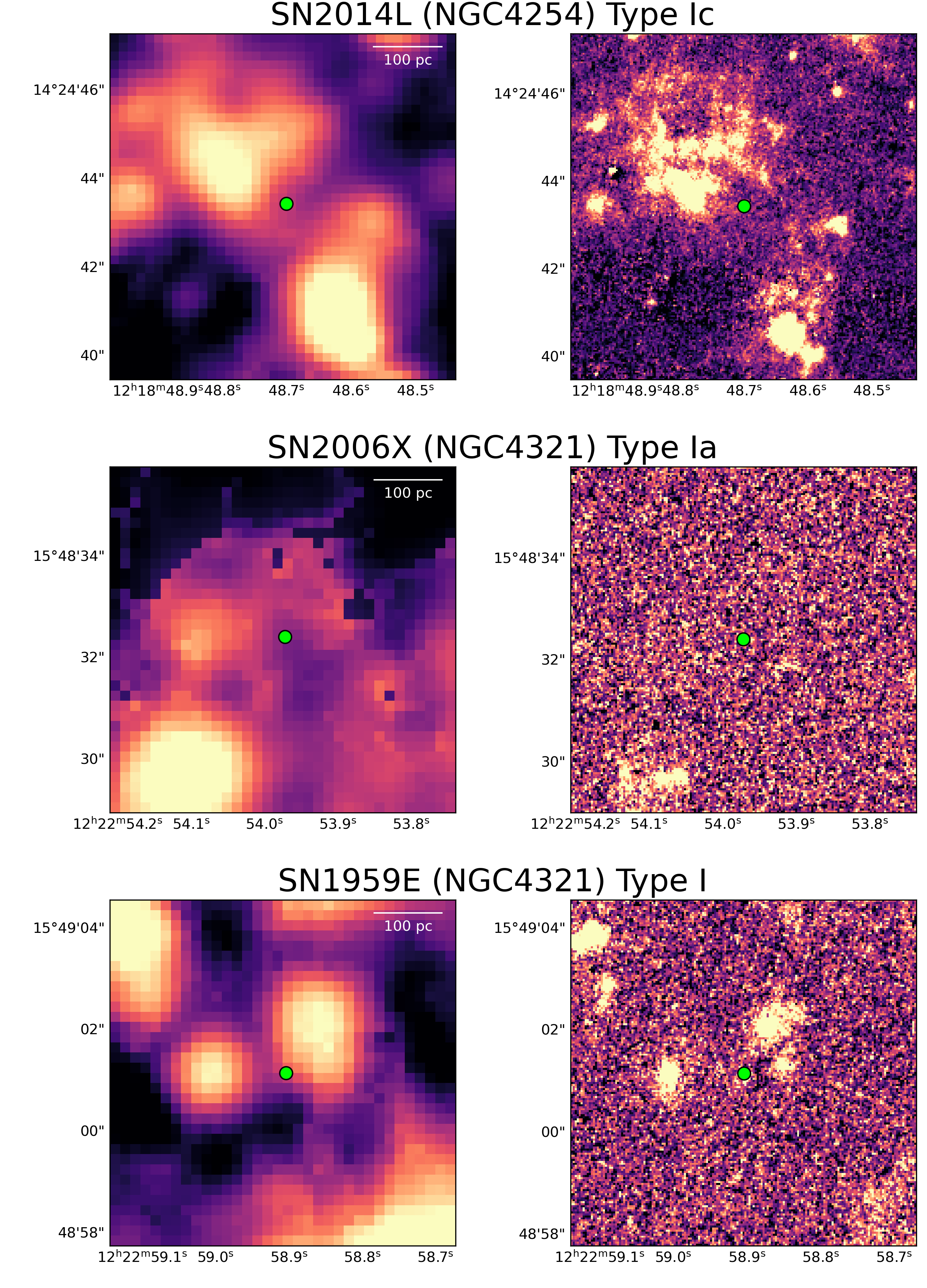}
\caption{Figure \ref{fig:HST} continued}
\end{figure*}
\begin{figure*}
\centering
\includegraphics[width=0.9\textwidth]{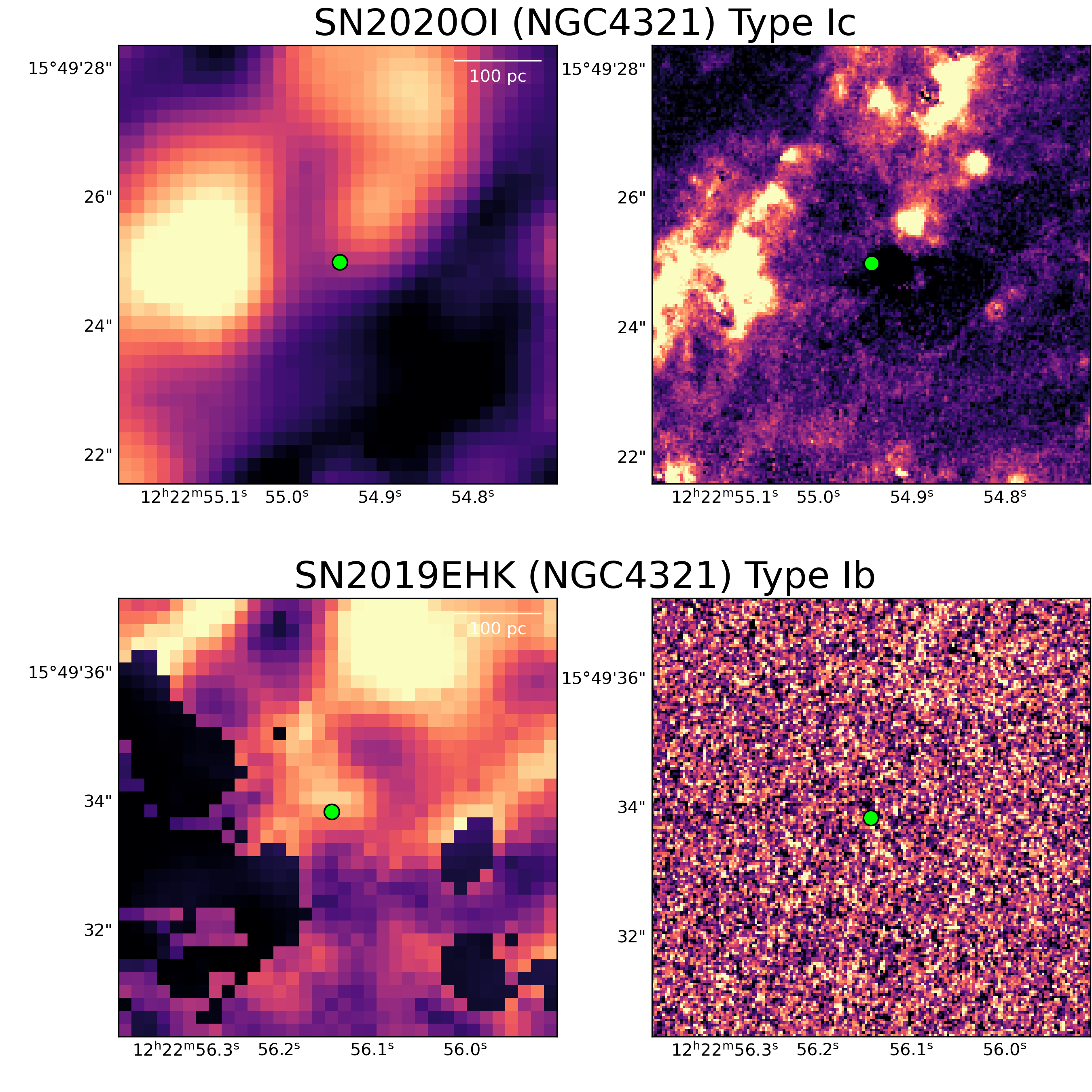}
\caption{Figure \ref{fig:HST} continued}
\end{figure*}

%%%%%%%%%%%%%%%%%%%%%%   APPENDIX   %%%%%%%%%%%%%%%%

\end{document}